\documentclass[sigconf,10pt,letterpaper]{acmart}

\AtBeginDocument{%
  \providecommand\BibTeX{{%
    \normalfont B\kern-0.5em{\scshape i\kern-0.25em b}\kern-0.8em\TeX}}}

\usepackage[english]{babel}
\usepackage{blindtext}
\usepackage[T1]{fontenc}
\usepackage{amsmath}
\usepackage{algorithmic}
\usepackage{url}
\usepackage{graphicx}
\usepackage{epstopdf}
\usepackage{enumerate}
\usepackage{subcaption}
\usepackage{paralist}
\usepackage[inline, shortlabels]{enumitem}
\usepackage{cleveref}
\usepackage{balance} 
\usepackage{xcolor}
\usepackage{mwe}

\crefformat{section}{\S#2#1#3} % see manual of cleveref, section 8.2.1
\crefformat{subsection}{\S#2#1#3}
\crefformat{subsubsection}{\S#2#1#3}

\newcommand{\ispa}{\textit{ISP A}\xspace}
\newcommand{\ispb}{\textit{ISP B}\xspace}

\acmConference[Manuscript]{To be decided}{}{}
\setcopyright{none}
\renewcommand\footnotetextcopyrightpermission[1]{}
\begin{document}

\title[Is Two Greater Than One?]{Is two greater than one?: Analyzing Multipath TCP over Dual-LTE in the Wild}

%\titlenote{Produces the permission block, and copyright information}
%\subtitle{Extended Abstract}

%\author{
%	%\alignauthor
%	\large Nitinder Mohan$^\dag$ \hspace{1em} Tanya Shreedhar$^\ddag$ \hspace{1em} Aleksandr Zavodovski$^\dag$  \hspace{1em}\\ Jussi Kangasharju$^\dag$ \hspace{1em} Sanjit K. Kaul$^\ddag$ \hspace{1em}
%	\\ 
%	\normalsize $^\dag$University of Helsinki, Finland\hspace{3em}
%	$^\ddag$IIIT Delhi, India  
%} 
\author[N. Mohan]{Nitinder Mohan}
\affiliation{
	\institution{University of Helsinki}
	\country{Finland}
}
\author[T. Shreedhar]{Tanya Shreedhar}
\affiliation{
	\institution{IIIT-Delhi}
	\country{India}
}
\author[A. Zavodovski]{Aleksandr Zavodovski}
\affiliation{
	\institution{University of Helsinki}
	\country{Finland}
}
\author[J. Kangasharju]{Jussi Kangasharju}
\affiliation{
	\institution{University of Helsinki}
	\country{Finland}
}
\author[S.K. Kaul]{Sanjit K. Kaul}
\affiliation{
	\institution{IIIT-Delhi}
	\country{India}
}
%\email{webmaster@marysville-ohio.com}
%\affiliation{%
%  \institution{Institute for Clarity in Documentation}
%  \streetaddress{P.O. Box 1212}
%\author{Lars Th{\o}rv{\"a}ld}
%\affiliation{%
%  \institution{The Th{\o}rv{\"a}ld Group}
%  \streetaddress{1 Th{\o}rv{\"a}ld Circle}
%\email{larst@affiliation.org}
%\affiliation{%
%  \institution{Inria Paris-Rocquencourt}
%\affiliation{%
% \institution{Rajiv Gandhi University}
%\affiliation{%
%  \institution{Tsinghua University}
%  \streetaddress{30 Shuangqing Rd}
%\affiliation{%
%  \institution{Palmer Research Laboratories}
%  \streetaddress{8600 Datapoint Drive}
%\affiliation{\institution{The Th{\o}rv{\"a}ld Group}}
%\email{jsmith@affiliation.org}
%\affiliation{\institution{The Kumquat Consortium}}
%\email{jpkumquat@consortium.net}
%\IEEEpeerreviewmaketitle
%\maketitle
%\IEEEpeerreviewmaketitle

\begin{abstract}
%Edge clouds are an attractive platform to support latency-sensitive applications by providing computations on servers deployed close to end-users. These servers aim to employ MPTCP to leverage multiple connections including wireless over a public network. In this paper, we show that the default MPTCP design does not adequately support reliability in these environments, which makes it unfit for use in edge clouds. We propose \sysname, an extension to MPTCP which focuses on adding reliability over network paths.  
Multipath TCP (MPTCP) is a standardized TCP extension which allows end-hosts to simultaneously exploit all of their network interfaces. 
%While it has so far seen most research in datacenter or heterogeneous wireless networks, 
The recent proliferation of dual-SIM mobile phones makes multi-LTE MPTCP setup an attractive option. 
We perform extensive measurements of MPTCP over
two LTE connections in low and high-speed mobility scenarios over five months, both in controlled and in-the-wild environments.
Our findings indicate that MPTCP performance decreases at high speeds due to increased frequency of signal
strength drops and handovers. 
Both LTE paths experience frequent changes
which result in a sub-optimal subflow utilization.
We also find that while path changes are unpredictable, their impact
on MPTCP follows a deterministic trend.
Finally, we show that both application traffic patterns and congestion control
variants impact MPTCP adaptability at high speeds.
% \textcolor{red}{Edit from here.} 
% Our findings indicate that (i) performance of
% MPTCP degrades as speed of the client increases, largely due to
% inability of scheduler and congestion control to adapt to variable path delays; (ii) coupled
% congestion control improves MPTCP's performance in different mobility, and (iii) application
% traffic patterns influences extent of adaptation to the changing network
% conditions. 
%Our study sheds light into MPTCP's behavior in all-mobile networks and is a basis for future cross-layer scheduling and congestion control work in this area.
\end{abstract}

\maketitle

\section{Introduction} \label{sec:intro}
%
%Internet's core protocols TCP and IP were designed for hosts connected via a single, wired network interface.
%%
%This state of affairs has changed drastically over the last decade due to the proliferation of smartphones, making cellular access to the Internet a very common case.
%%Since the inception of Transmission Control Protocol (TCP) and Internet Protocol (IP), the state of networking today has changed considerably. 
%%
%%Internet protocols were designed considering a simplistic networking view where interacting hosts are connected via a single (often wired) network interface.
%%
%%In current times, there has been significant advancements in aspects of scale, reliability and usability of the Internet. 
%%
%%Where the number of connected hosts have seen a rapid increase due to proliferation of smartphones, the preferred means of accessing the Internet has now become cellular.
%
According to CISCO, global mobile data traffic has grown \textit{17-fold} between 2012-2017 and 71\% in a single year~\cite{ciscowhitepaper}.
%
%Latest cellular technologies such as LTE promise mobile users high speed, high bandwidth connectivity at low costs.
%
LTE is widely deployed \cite{ltecoverage} and future technologies, such as 5G, strengthen the need for efficient protocols over cellular links.
%LTE deployments already have a very good coverage~\cite{ltecoverage} and future technologies, such as 5G, are likely to further strengthen the need for efficient protocols over cellular links.
%The case of LTE is supplemented by extensive global adoptability by Internet Service Providers (ISPs) leading to a substantial availability of network \cite{ltecoverage}.
%
%Studies have been conducted to understand the performance of TCP over LTE~\cite{huang2012, nguyen2014}.
%Despite theoretical advantages,
Studies show that TCP performs poorly over LTE, especially when the user is mobile, primarily due to the large variability in network conditions~\cite{huang2013, nguyen2014}.
%
%The key takeaway is that TCP throughput degrades significantly over cellular networks, especially when the user is mobile, caused primarily by rapidly varying network conditions and outages due to handovers. 
%
This deterioration is further heightened due to presence of large buffers within the ISP network which often results in TCP connection stalls and ineffective link utilization due to excessive queueing~\cite{huang2013, jiang2012}.
%To mask link layer disruptions from user applications, ISPs employ large packet buffers which often cause significantly varying packet delivery rates, connection stalls and reduction in effective bandwidth utilization~\cite{huang2013}.

Multipath TCP (MPTCP) is a TCP extension allowing unmodified applications to leverage multiple network interfaces to form parallel TCP connections between end-hosts~\cite{mptcplinux}.
%%
%Although it was originally designed for datacenter networks~\cite{raiciu2011}, MPTCP's inherent ability to aggregate bandwidth over multiple paths has also been leveraged in mobile environments~\cite{mptcpmobile}.
%
%Due to its added benefits of reliability and robustness over heterogeneous wireless paths such as WiFi, LTE etc., r
%Due to its benefits, researchers have proposed MPTCP as a transport protocol for several emerging applications and technologies such as augmented reality~\cite{mptcp_ar}, vehicular networks~\cite{raven}, edge clouds~\cite{mohan2018}, Internet-of-Things (IoT)~\cite{energymptcp}, etc.
%
Researchers have studied and analyzed its impact on emerging technologies (AR/VR~\cite{mptcp_ar}, IoT~\cite{energymptcp}, edge computing~\cite{mohan2018} etc.) while exploiting multiple network types such as WiFi, cellular and ethernet~\cite{deng2014wifi}.
MPTCP kernel is available as open-source and is in use, e.g., by Apple in their iOS devices~\cite{mptcpapple}.
Despite these efforts, MPTCP faces several challenges in mobile networks, such as failing to use heterogeneous network combinations with large delay differences, e.g. WiFi and LTE~\cite{blest}.
%
%In such heterogeneous networks, MPTCP experiences additional reordering delays 
%due to reordering out-of-order packets from different paths 
%at the receiver which can limit overall bandwidth utilization, often plummeting it worse than a single TCP.  
%Previous studies of MPTCP over wireless have primarily focused on heterogeneous networks, i.e., WiFi and cellular which exhibit significant end-to-end delay differences \cite{deng2014wifi, blest}.  
%
%MPTCP usage in such environments is negatively impacted by high delay differences between connections leading to under-utilization of available aggregate bandwidth \cite{}. 
%Previous studies of MPTCP over wireless have primarily focused on heterogeneous networks, i.e. WiFi and cellular with significant path delay differences~\cite{blest, deng2014wifi}.
%
However, we believe that MPTCP's most pragmatic use is in multi-LTE networks for two reasons. 
\emph{First}, modern smartphones are often equipped with dual-SIM slots and chipsets enabling the use of two LTE connections.~\cite{dsdspatent}.
%
%Secondly, large coverage areas and high bandwidth offered by multiple LTE connections can enable effective deployment of applications/technologies relying on MPTCP as their core transport protocol.
\emph{Second}, vast coverage areas, large combined bandwidth and reliable packet delivery offered by multi-LTE connections can enable effective deployment of emerging technologies discussed above.
Also, intuitively, none of MPTCP heterogeneity issues should affect a multi-LTE setup as both paths exhibit similar characteristics with similar delays to the server. 

In this paper, we conduct to our knowledge the first comprehensive measurement study of MPTCP over multi-carrier LTE connections in \emph{day-to-day mobility} scenarios.
%investigate the performance and behavior of MPTCP in multi-carrier LTE networks. 
%
%Unlike previous studies which have either considered static~\cite{mptcpmore} or very high-speed mobility (>250~km/h)~\cite{li2018}, we focus our work on day-to-day mobility scenarios, i.e. walking, driving, etc.
%
%We conduct a large scale measurement study with a client using LTE connections from two major Finnish cellular providers, connected to a server hosted in a remote cloud.
% conventional network of mobile client being equipped with LTE connections from two major cellular carrier providers based in Finland and server based in a remote cloud.
%
%\textcolor{red}{Write other info about the study here. Delete sentences below. Section below requires editing.}
%
Over a period of \textit{five months}, we performed extensive data collection in controlled environments and in-the-wild to understand the impact of last-mile quality, mobility and application workload on MPTCP performance.
%
%We compare the effect of changes in network quality, notably signal strength drops and handovers, to MPTCP performance for different application loads and network parameters.
%
%To the best our knowledge, this is the first systematic analysis of MPTCP behavior with physical network changes due to typical mobility over multi-carrier LTE networks.
%
Our key findings are:% as follows.
\begin{enumerate}[wide, labelwidth=!, labelindent=0pt, topsep=0pt]

\item MPTCP bandwidth gains over two LTE connections decrease significantly with increasing mobility.
Downloads over MPTCP take over twice as long at speeds >60km/h, performing even worse than a single TCP most of the time.

\item Frequent signal strength drops and handovers are the cause of MPTCP deterioration at high speeds.
Such changes can induce 10-fold RTT spikes on a subflow resulting in severely increased queuing and reordering delays.  

\item Contrary to our intuitions, we find that MPTCP exhibits a skewed subflow utilization as both LTE paths experiences last-mile changes at varying time and frequency which often overlaps with each other.
%Contrary to our intuitions, we find that both LTE paths experience last-mile changes at varying time and frequency, often overlapping with each other which results in a skewed subflow utilization in MPTCP. 

\item Although LTE last-mile changes are unpredictable 
%and equally common on both LTE connections 
at high speeds, their impact on MPTCP follows a deterministic trend.
This allows monitoring occurrences of link changes and circumventing their effects to maintain MPTCP's benefit over multiple connections. 

\item Tweaking application traffic burstiness and congestion control schemes can also assist the protocol in its adaptability and robustness at high speeds. 
Our analysis shows an improvement of 75\% QoE and 18\% throughput in MPTCP. 

\end{enumerate}

%Include if available space
%\noindent \textbf{Paper Organization.}
%%
%After discussing the background and related work in \cref{sec:background} we present our measurement methodology in \cref{sec:measurement}.
%%
%We further discuss our findings from controlled mobility tests in \cref{sec:controlled}  and analysis from uncontrolled data collection in \cref{sec:uncontrolled} .
%%
%We explore several possible future works in \cref{sec:discussion} before concluding our paper in \cref{sec:conclusion}.
%%% Local Variables:
%%% mode: latex
%%% TeX-master: "paper"
%%% End:

%
\section{Background \& Related Work} \label{sec:background}

MPTCP is in-kernel extension to TCP that allow multi-homed hosts to use multiple parallel TCP connections over each network interface. 
Data packets from the application are scheduled to one of the underlying TCP subflows by \emph{scheduler}~\cite{shreedhar2018qaware}.
The default \texttt{minSRTT} scheduler~\cite{minsrtt} prioritizes the subflow with lowest smoothed round trip time (SRTT) to the receiver.
%
%In highly variable delay network environments, the default scheduler switches between subflows quite frequently leading to an increased out-of-order transmissions~\cite{sinky2016}.
%Using minSRTT scheduler in variable delay networks, leads to packets arriving out-of-order at the receiver due to packets experiencing different delays on the paths ~\cite{sinky2016}.
In variable delay networks, the scheduler is known to produce
out-of-order packets at the receiver as they experience different delays on parallel paths~\cite{sinky2016}. 
%the packets encountering different delays on paths arrive out-of-order at the receiver~\cite{sinky2016}. 
%The default scheduler, \textit{minSRTT}~\cite{minsrtt}, schedules packets to path with minimum smoothed round trip time (SRTT), provided the subflow has available space in its congestion window.
%
%The paths can have different delay characteristics and it is the task of the scheduler to decide where to send the next data-byte. 
%Like TCP, congestion control algorithm is an important aspect of MPTCP. This allows it to dynamically adapt to changing network conditions on all the available paths.
%
%The receiver waits and reorders all incoming packets in its receive buffer and delivers them in-order to the application.
Such packets are buffered at the receiver and are only delivered "in-order" to the application.
MPTCP minimizes additional reordering delays by employing \emph{coupled congestion} mechanisms at the sender which balance packet congestion over all subflows~\cite{raiciu2012}.
%

%MPTCP employs coupled congestion control algorithms
%, in addition to TCP congestion schemes, 
%which aim to balance congestion between all underlying TCP flows~\cite{raiciu2012}.
%The main goals of congestion control is to ensure fairness to TCP at the bottleneck link, MPTCP should perform atleast as well as the regular TCP on the best path along with scheduling more data on less congested paths .  This is done by maintaining per subflow congestion window which is halved in case of packet losses, as in standard TCP. However, the total increase in congestion windows is such that it is fair to single path TCP.
%
%It has been observed that MPTCP performance is limited by its receive window reordering in scenarios where individual subflows have high delay differences~\cite{sinky2016}.
%
%due to receiver buffer blocking where subflows exhibit high delay differences~\cite{blest}.
Previous studies have focused on analyzing MPTCP over heterogeneous networks such as WiFi and cellular.  
The key takeaway is that MPTCP performance is limited due to consistent delay difference between both paths resulting in ineffective utilization~\cite{deng2014wifi, nikraveshmptcpmobile}.
Little attention has been paid to MPTCP in multi-LTE networks.
% which exhibit a unique behavior.
%
LTE connections are resilient to packet losses and offer high bandwidth aggregation opportunity for MPTCP~\cite{huang2013}.
However, studies analyzing TCP over LTE have shown that with increasing mobility, TCP performance degrades due to delays in connection establishment, timeouts and interruptions~\cite{nguyen2014, li2016}.
%
%Furthermore, frequent handovers on the network can also result in large packet transmission delays.
%
%\citet{tcpConcise2014} and \citet{xu2017} 
At high speeds, TCP packets experience excessive on-device and in-network queuing along with packet losses \cite{tcpConcise2014, xu2017}.
Therefore, it is pertinent to understand whether MPTCP provides any benefits to mobile clients while utilizing multiple LTE paths. 

Study by \citet{li2018} is closest to ours. 
While the authors analyze MPTCP using dual-LTE in very high-speed rails
(>250km/h), our work focuses on understanding MPTCP's behavior in day-to-day mobility using generic transportation modes (<100km/h), such as walking, driving, etc.
Furthermore, the focal point of authors work in \cite{li2018} was to study LTE handover's impact on MPTCP performance.
On the other hand, we investigate the correlation between any/all last-mile link changes (signal drops, handovers) and other network parameters (app traffic, congestion control) on MPTCP. 
%Little to no attention has been paid to understanding MPTCP behavior over multiple LTE connections in day-to-day mobility \textcolor{red}{signal}. 
   
%In this study, we focus on understanding MPTCP behavior over multiple LTE connections in low to high mobility speeds.
%while mobile due to signal drops and handovers. 
%
%Our goal is to analyze whether MPTCP is capable of utilizing \textit{similar} physical networks in fluctuating conditions. 

%We study the effect of mobility, congestion control and application load on MPTCP performance.
 
%%% Local Variables:
%%% mode: latex
%%% TeX-master: "paper"
%%% End:

%
%\section{Measurement Setup \& Data Collection} \label{sec:measurement}
\section{Measurement \& Analysis Method} \label{sec:measurement}
% \begin{figure}[!tb]
%     \centering
%     \includegraphics[width=0.4\linewidth]{Figs/rpi_photo}
%     \caption{Mobile Client: Raspberry Pi 2 equipped with two LTE modems and a battery.}
%     \label{fig:client}
%     %\vskip -3mm
% \end{figure}

%The focus of our study departs from previous work in fundamental way, which was key to our experiment design and methodology. While past studies have analyzed MPTCP over multiple LTE connections in respect of bandwidth aggregation, we focus on understanding cross-layer correlation between network changes on LTE to MPTCP performance. 
%We describe our measurement methodology, design of our setup and parameters collected in our measurements. 

Our aim is to \emph{analyze MPTCP performance over two distinct LTE connections in different mobility}.
We elected to use a device equipped with two separate LTE antennas.
Current dual-SIM enabled smartphones are fitted with a single antenna which is time-shared by both connections~\cite{dsdspatent}.
Our aim in using two antennas is to establish the upper bound on MPTCP
performance and not be limited, e.g., by NIC queuing due to resource contention.
%
%This imposes several constraints on our setup.
%
% \emph{First}, our test device needs to be equipped with two separate LTE antennas to support parallel transmissions.
% %
% Current dual-SIM enabled smartphones are fitted with a single antenna which is time-shared by both connections~\cite{dsdspatent}.
% %
% Although this does not impact end-user's QoE, lack of parallel antennas will force MPTCP to utilize one path at a time and may result in excessive NIC queuing due to resource contention.
%
We also wanted to use multiple ISPs since the internal network configuration, such as scheduling, routing, priority queue etc., can differ for each ISP depending on its QoS promises.  
%every ISP configures its network differently which best suits the QoE it promises its subscribers, e.g. scheduling algorithms, routing, priority queues etc.   
%
%Restricting data measurements to a single provider can introduce an unwanted bias in our results.
%The design of our setup, shown in Figure \ref{fig:sysArch}, centers on analyzing cross-layer correlation between physical network changes on LTE to MPTCP. 

\subsection{Setup Configuration}

\begin{figure}[!t]
    \centering
    \includegraphics[width=0.98\linewidth]{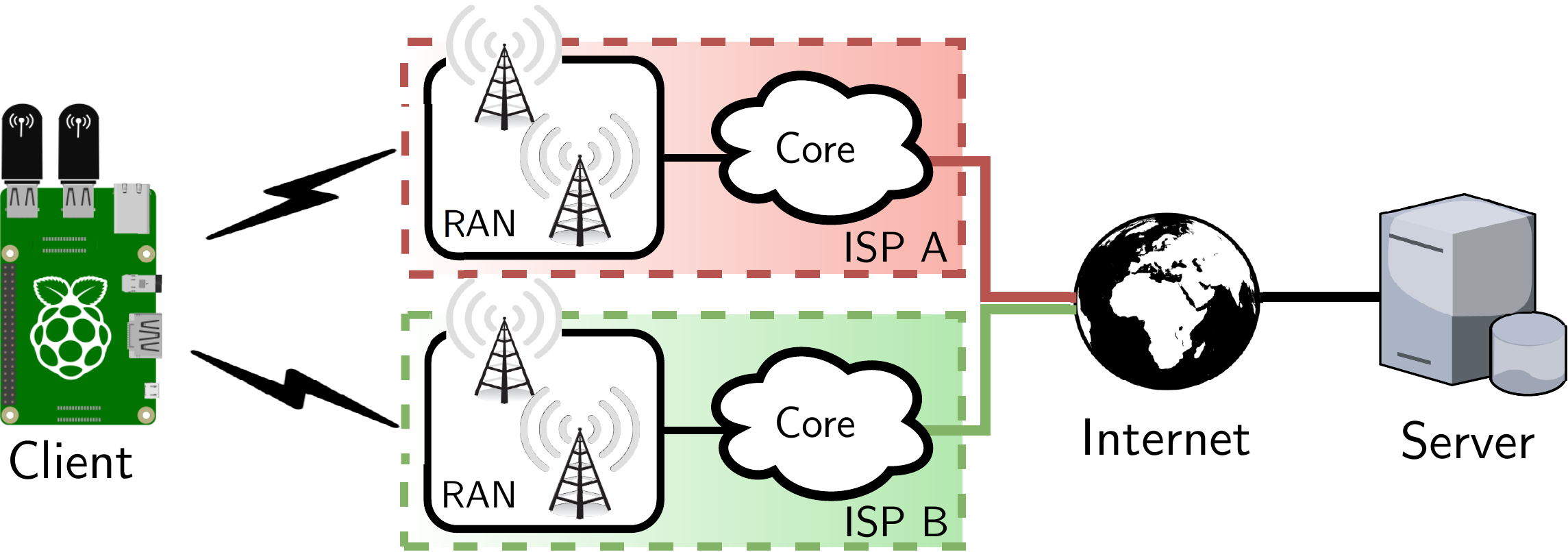}
    \caption{Data collection setup. Client is MPTCP-capable RPi equipped with two LTE connections from different ISPs. Server is an AWS ec2 instance.}
    \label{fig:sysArch}
    \vspace*{-1em}
\end{figure}

We designed our measurement setup as shown in \autoref{fig:sysArch}.
% illustrates our measurement setup. 
%
We conducted our experiments in the capital region of a European country from \textit{September 2018} to \textit{February 2019}.
%\noindent \textbf{Client.} 
Our test device is a Raspberry Pi 2 (RPi) equipped with two Telewell CAT4 LTE USB modems.
The RPi runs Raspbian OS over latest MPTCP v0.94~\cite{mptcplinux} and is powered by an external battery to enable mobility.
We equip USB modems with LTE connections from \emph{two} major cellular providers, \ispa and \ispb; capped to 150~Mbps downlink and 50~Mbps uplink.
Both ISP's offer extensive network coverage in the region with \emph{almost equal} LTE basestation (BS) deployment density.
%and non-intersecting RAN infrastructure.
As the server, we set up AWS ec2 instances running MPTCP v0.94 on 32 GB RAM, 1 Gbps ethernet, 16-core 2.4GHz CPU and Ubuntu 18.04. 
Both cellular providers have \emph{non-intersecting} network path to AWS, which we verify by periodically running \texttt{traceroute} over both connections. 
%Both LTE connections are limited to 150~Mbps downlink and 50~Mbps uplink.
%
%The client runs Raspbian OS over latest MPTCP kernel v0.94 based on Linux kernel v4.14~\cite{mptcplinux}.
%
%Unless otherwise specified, the default configuration of MPTCP utilizes \textit{minSRTT} scheduler and uncoupled CUBIC congestion control~\cite{tcpcubic}. 
%
%The sizes of both send and receive buffers were set at 4MB (which is big enough for our network setting).
%
%Finally, the entire module is made mobile by powering the RPi through an external battery pack.
%\smallskip
%We automate our experiments using several Python and bash scripts.
%
%\noindent \textbf{Server.}
%
By default, both client and server use \textit{minSRTT} scheduler and CUBIC congestion control~\cite{tcpcubic}.
%In our initial tests, we confirm server's capability to support two simultaneous LTE transfers without acting as a bottleneck.
%
In stable conditions, we observe $\approx54$~ms RTT to the server over both ISP connections.

%\begin{figure}[!t]
%	\centering
%	\includegraphics[width=0.45\linewidth]{Figs/drive_heatmap_new}
%	\caption{LTE base stations of both ISPs along driving path}
%	\label{fig:driveHeatmap}
%	%\vskip -3mm
%\end{figure}

%\begin{figure}
%  \begin{minipage}[b]{0.49\linewidth}
%     \includegraphics[width=\linewidth]{Figs/rpi_photo}
%     \caption{Raspberry Pi client with LTE modems \& battery.}
%     \label{fig:client}
%  \end{minipage}%
%  \hfill
%  \begin{minipage}[b]{0.47\linewidth}
%    \includegraphics[width=\linewidth]{Figs/drive_heatmap_new}
%    \caption{Base stations of the ISPs along driving route.}
%	\label{fig:driveHeatmap}
%  \end{minipage}
%\end{figure}

\subsection{Data Collection}
We built a BASH-based \textbf{data collector} for RPi which performs the following tasks: \emph{network measurements and mobility detection}, \emph{data transfer} and \emph{upload to measurement server}.

\vspace{-0.5em}
\subsubsection*{\textbf{Network Measurements.}}\label{subsec:netMeasurements}
Every second (defined as "period") the data collector queries attached LTE modems with AT
commands and logs their current signal strength (dBm) and associated
basestation ID (\textit{BSID}). % to local database.
As we wish to analyze the impact of mobility on MPTCP, it is necessary to detect when the RPi is mobile.
However, unlike smartphones, the RPi cannot infer mobility using conventional techniques due to lack of a built-in accelerometer or GPS.
We overcame this limitation by constantly monitoring changes in signal strength ($\delta_{signal}$) over both modems. %\emph{between periods}.
The collector flags $\delta_{signal} \geq$ 6dBm on either modem as ``mobility start" and initiates data transfer.
% 
%Our choice of 6dBm mobility threshold value is inspired from our controlled measurements (\cref{sec:controlled}) 
%We found 6dBm to be best fit for mobility threshold value based on our experiments in controlled settings (\cref{sec:controlled}). 
We found 6dBm to be the best fit for mobility threshold as it achieves the maximum accuracy (98.3\%) with zero false negatives in contrast to other values in our controlled experiments (\cref{sec:controlled}).
Also, we denote change in BSID between consecutive periods as successful handover.

\vspace{-0.5em}
\subsubsection*{\textbf{Data Transfer \& Measurement Upload.}} \label{subsec:dataTransfer}
We developed a client-side application which downloads a 300MB\footnote{We opt for a longer flow as MPTCP is known to perform badly for short-lived flows due to slow congestion window growth~\cite{nikraveshmptcpmobile}.} file from the server over HTTP.\footnote{We leave measurement of uplink performance as future work.}
We also conducted DASH video streaming measurements over MPTCP to provide further granularity to our analysis (\cref{subsec:video}).
During data transfer, the collector passively records packet traces via \texttt{tcpdump} which we use to analyze several MPTCP/TCP metrics such as RTT, bytes-in-flight, out-of-order queue size etc.
%
%As MPTCP is known to perform badly for short data transfers~\cite{nikraveshmptcpmobile}, we restrict our download size to 300 MB.
%
%
The collected data in the RPi is uploaded to our measurement server every night if $\delta_{signal}$ remains zero for two consecutive hours.
     
%1. Details of controlled and uncontrolled collections, i.e. gave devices to volunteers 2. what parameters we collect and how we estimate speed from network changes. 3. Scripts used to collect data.
%Throughout this study, we empirically demonstrate the impact of LTE wireless access changes on scheduling and congestion mechanisms of MPTCP.
%%
%We test different TCP applications such as HTTP file download, video streaming etc., in different mobility conditions and examine different congestion control schemes designed for MPTCP, namely Linked Increase Adaptation (LIA)~\cite{lia}, Optimized Linked Increase Adaptation (OLIA)~\cite{olia} and Balanced Link Adaptation (BALIA)~\cite{balia}.
%%
%We complement our data traces, captured via \texttt{tcpdump}, with current signal strength (in dBm) and base station ID (\textit{CellID}) for each LTE connection logged at per-second precision.
%%
%We record these network quality metrics through a specifically designed bash script which queries attached LTE modems with AT commands~\cite{atcommand} and logs the output to a database.
%%
%Changes in \textit{CellID} can be inferred as handovers on a particular subflow, however this method only detects completion of a handover but not when it started.
%%
%We compensate for this imprecision by marking a \textit{one} second window over each \textit{CellID} change as ``handover'' event.
% which allows us to incorporate possibilities of the connection experiencing either lossless or seamless handover.

\begin{table}[!t]
\small
\begin{tabular}{c|cccc}
                      & \begin{tabular}[c]{@{}c@{}}MPTCP \\ Conns. \end{tabular} & \begin{tabular}[c]{@{}c@{}}TCP \\ Conns.\end{tabular} & \begin{tabular}[c]{@{}c@{}}File\\ DLs\end{tabular} & \begin{tabular}[c]{@{}c@{}}Video\\ plays\end{tabular} \\ \hline
\textbf{Uncontrolled} & 10863                                                   & 7966                                                  & 7902                                               & 1793                                                  \\
\textbf{Controlled}   & 1345                                                    & 937                                                   & 825                                                & 192                                                  
\end{tabular}
\caption{Measurement statistics.}
\label{table:measurementStatistics}
\vspace*{-2em}
\end{table}

%\footnotetext{Terms "connection", "flow" and "subflow" are used interchangably.}
We gave our test devices to three volunteers to carry along their daily commute, encompassing walking, driving and public transport such as trams, buses etc. (see \cref{sec:uncontrolled}).
We did not record any real-world information such as distance traveled, locations, mode of transport etc.
We also conducted multiple measurements in controlled environments where we chart out a planned test route for different mobility speeds (see \cref{sec:controlled}).
%
%We parallelly perform experiments over two individual TCP flows mapped to each ISP connection. 
%This allows us to analyze and compare MPTCP subflow behavior to regular TCP experiencing same network changes at the same time.
Controlled experiments serve us two purposes;
(1) developing classification models over network changes for grouping our in-the-wild traces into different mobility categories; 
and, (2) closer inspection of MPTCP behavior.
\autoref{table:measurementStatistics} shows our measurement statistics.  
%
%Overall, our collection lasted from \textit{September 2018} to \textit{February 2019} and resulted in $\approx40$~GB of data. 
%
%We plan to release our dataset for use by the research community in near future. 

%%% Local Variables:
%%% mode: latex
%%% TeX-master: "paper"
%%% End:

%
\section{Impact of LTE Mobility on MPTCP} \label{sec: results}

%In this section, we study the effect of cellular mobility on overall performance of MPTCP.
%
We first analyze MPTCP's performance in controlled mobility tests and provide further explanation to trends in its behavior by scrutinizing the data collected in-the-wild.
%   
%Instead of seperate sections for controlled and uncontrolled, we will have a single unifying section where we will put all results.
%
%The structure is going to be as follows:

\subsection{Controlled Mobility Measurements} \label{sec:controlled}
%
%We first explain our results from controlled measurements. We start with a general description of the test route followed by what we want to discuss in this section. We can directly refer to the result table and then show surprise in why MPTCP starts to perform so bad. 

\begin{table}[!t]
\small
\begin{tabular}{c|cc|cc|cc}
                      & \multicolumn{2}{c|}{\textbf{Rate (Mbps)}} & \multicolumn{2}{c|}{\textbf{RTT (ms)}} & \multicolumn{2}{c}{\textbf{DL Time (s)}} \\
 \textbf{Mobility} 	& MPTCP             & TCP             & MPTCP          & TCP          & MPTCP           & TCP           \\ \hline
Stationary            & 69.2                & 42.1              & 56             & 57           & 36.6              & 71.4            \\
Walking               & 42.5                & 34.8              & 65             & 68           & 61.2              & 78.1           \\
Driving               & 31.9               & 31.7              & 74             & 75           & 81.5 		& 82.3           
\end{tabular}
\caption{Overview of controlled measurement results.}
	\label{table:control}
	\vspace*{-2em}
\end{table}

We conducted multiple measurements for two common mobility types,
\emph{walking} and \emph{driving}, using a fixed route for both modes.
Our driving route was 13km long inter-city highway while our walking path was 4.7km encircling the city center. 
We maintain an average speed of $\approx6$km/h and $\approx80$km/h while walking and driving, respectively.
The RPi was placed near the vehicle's windshield to avoid any signal shielding by the chassis. %while driving.    
While we regulate the speed in our experiments, we have no control over the underlying network environment.
We eliminated any outlier bias by performing multiple iterations of each experiment at different times of the day.
We also performed \emph{baseline experiments} where the RPi was kept stationary throughout the data transfer.

\autoref{table:control} compares the average throughput, RTT and file download time obtained by MPTCP and TCP over LTE from each ISP. 
%
%It is made evident from the results that \emph{increasing mobility significantly degrades MPTCP performance in LTE networks}.
%
%\textcolor{red}{First describe the MPTCP results as in percent decreases. Then talk about DL time vs throughput.}
Both MPTCP and TCP show a decline in performance with increasing speeds.
%linearly decline trend in performance with increasing speeds over MPTCP.
%
Although the RTTs for both protocols show a similar degrading trend, the impact of high speed on throughput and file downloads is more significant in MPTCP.
While TCP exhibits 10\% reduction in throughput and download time, MPTCP throughput decreases by 37\% and downloads take \emph{twice} as long while driving.
Interestingly, MPTCP's bandwidth gains over two LTE connections lessen at higher speeds as it achieves similar throughput as TCP. 
%
%To understand this degradation, we dissect our collected traces further.
We investigate the root cause of this degradation by dissecting our collected traces.
%While walking, MPTCP throughput decreased by $\approx25\%$ decrease in throughput and download time, and $\approx15\%$ increase in RTT and packet drops (not shown).   
%
%MPTCP throughput reduced by $37\%$ in our driving tests, but file download took almost \emph{double} time compared to a stationary client.
%
%On further analysis, we find that the increase is partly due to rise in packet drops (34\%) often leading to delays in HTTP connection establishment.
%
%These problems did not impact TCP as significantly, which only exhibited 10\% decrease in performance while driving.  
%On the other hand, the impact on TCP is not as significant compared to MPTCP showing performance degradation of .
%
%Surprisingly, we observe that MPTCP over two LTE connections achieves similar performance as a single TCP LTE at high speeds.
%  
%To uncover the root cause of this deterioation, we dissect our collected data traces.

\begin{figure}[!tb]
	\centering
	\includegraphics[width=\linewidth]{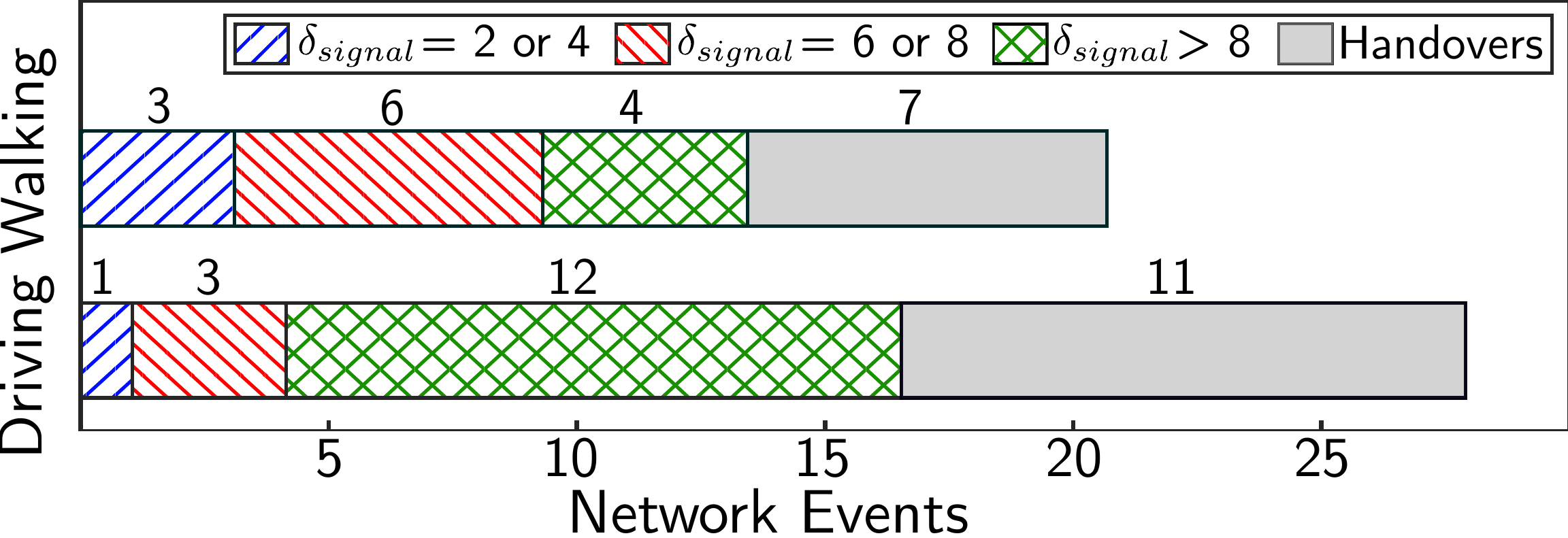}
	\caption{Network event frequency every five minutes in dual-LTE in controlled experiments.}
	\label{fig:probChange}
	\vspace*{-1em}
\end{figure}

%\noindent \textbf{Network quality to mobility signature.} Essentially we explain the number of network changes observed with mobility and what kind of signature we get.
%%
%This signature can be used in classifying and sorting the collected measurements in uncontrolled analysis.

\vspace{-0.5em}
\subsubsection*{\textbf{LTE link changes with mobility.}} \label{subsec:controlledLastMile}
As the mobility only changes the last-mile link (USB modem$\leftrightarrow$LTE basestation),
%As we only change the last-mile link (USB modem$\leftrightarrow$LTE basestation) by moving RPi in our experiments, 
%As throughout our experiment the only change in the network is at the last-mile, 
the first question we answer is, \emph{what is the impact of client mobility on last-mile LTE?}
%start by quantifying different physical link changes observed in our experiments. 
%
An LTE connection is likely to experience both changes in signal strength and handovers while the client moves closer/away from the BS. 
Our analysis shows that only \emph{drops in signal strength} ($\delta_{signal}$) and \emph{handovers}\footnote{Collectively referred as \emph{network events} throughout the paper.} have an impact on MPTCP throughput.    
%
%To study the effect further, 
%we segment network quality data recorded in each mobility experiment into \textit{5 minute} durations for both connections and 
We quantized occurrences of all network events on both LTE connections while driving and walking.
Figure~\ref{fig:probChange} shows the frequency of different network events every five minutes on both connections.
%number of low to large signal drops, (2-8~dB and $>8$~dB) and handovers observed within data transfer segments. 
We observed a 35\% increase in overall events along with a 57\% increase in handovers while driving compared to walking.
%Unsurprisingly, we observe that the client experiences 35\% increase in overall network state changes between walking and driving and 57\% increase in handovers.
%
Furthermore, while drops $\leq8$dBm are more frequent for a walking client (69\%), signal drops >8dBm predominate network events observed while driving (75\%).
%are predominated by low-intensity signal drops upto 8dBm (69\%) while drops greater than 8dBm were more prominent at higher speeds (75\%).  
% We see that the number of handovers observed by the client during driving far exceeds that while walking.
%Although a handover event is likely accompanied by a drop in signal strength, this was not found true in all cases. 
%Even though we observed significantly more handovers in driving tests, 
We also find that 43\% and 52\% handovers follow a signal drop while walking and driving, respectively.
%
%However, the numbers are likely influenced by design of our collector which records network parameters at per second granularity.    
%However, we also notice a peculiar \textit{"ping-pong"} effect in our walking measurements where the client frequently switches between the same set of base stations without experiencing any drop in signal strength.
%
%Future measurements using sensors with greater precision may improve these results further.
% our walking frequent handovers in short time intervals in the latter.
%
%In several segments, we 
%
%Although we were unable to find the root cause of this effect, we strongly suspect this behavior to workings of the basestation load balancing algorithms employed by the ISP in areas with denser deployment\textcolor{red}{[ref?]}.
% 
%Furthermore, our results show that a walking client is likely to experience signal drops of $\approx8$~dBm whereas the intensities of the drops while driving are much higher.
%while walking a client is more prone to see up to 8db signal strength drops on either LTE connection which is enough to trigger a lower MCS and PHY data rate.
%
%As we will show later, this can adversely affect the overall performance of MPTCP.
%
%On the other hand, we observed significant signal strength drops coupled with handovers while driving which lead to a notable degradation in MPTCP performance (shown in ).

\noindent \underline{\texttt{Takeaway1}:} %Increase in speeds leads to an increased frequency of network changes on last-mile LTE. 
%
%MPTCP over dual-LTE can experience 57\% increase in handovers and and $2\times$ high-intensity signal drops while driving in contrast to walking.
%\textcolor{red}{write in context of MPTCP}.
%Increasing mobility result in increasing last-mile link changes which are almost equally distributed over both LTE connection 
\textit{Both LTE connections experience increased frequency of network events at the last-mile with increasing mobility.
While low-intensity signal drops are more probable for slow-moving clients, the rate of handovers and large signal drops increases for clients moving at higher speeds.}
%LTE encounters more handovers and large signal drops at high speeds. 

\begin{figure}[!t]
\centering
\begin{subfigure}{0.235\textwidth}
\centering
  \includegraphics[width=\textwidth]{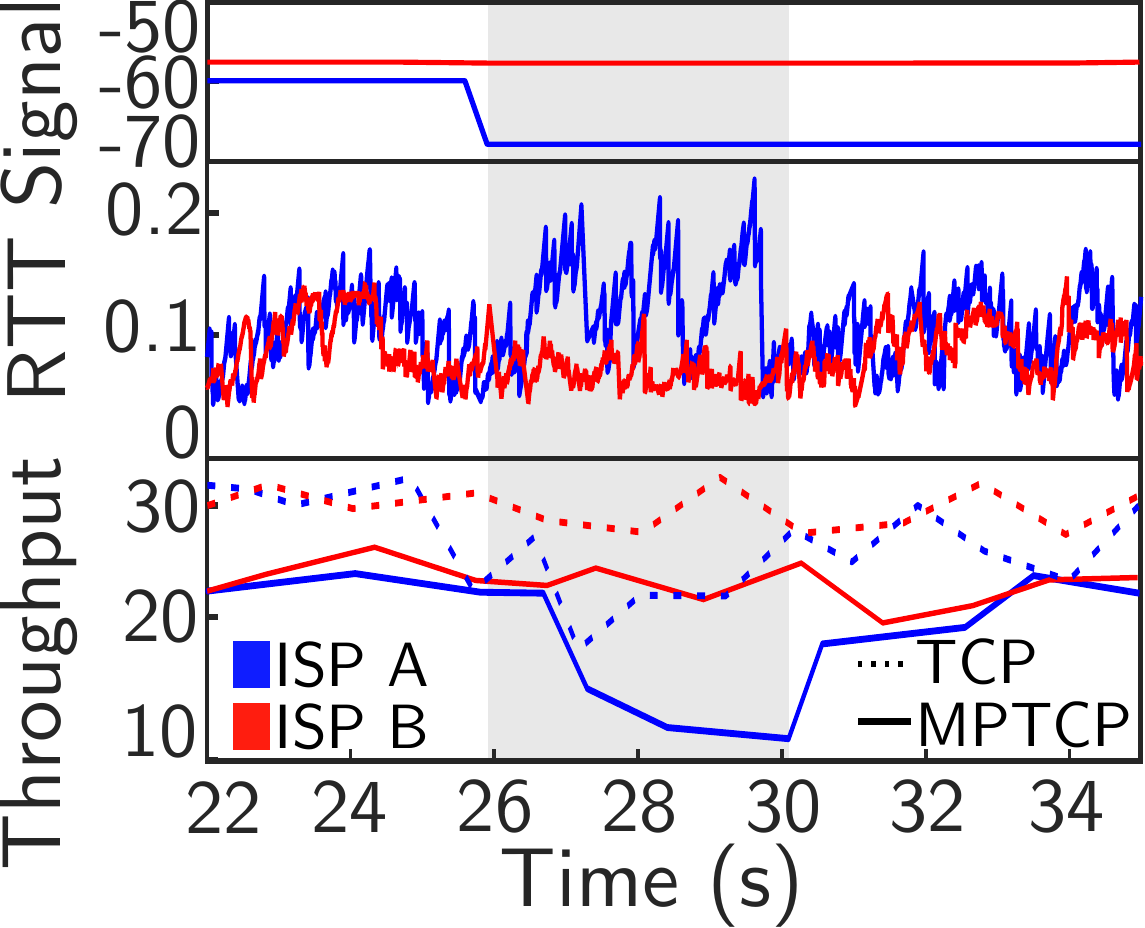}
  \caption{\label{fig:lineSNR}Signal strength drop}
\end{subfigure}%
%\hfill
\hspace{0.001\textwidth}
\begin{subfigure}{0.235\textwidth}
\centering
  \includegraphics[width=\textwidth]{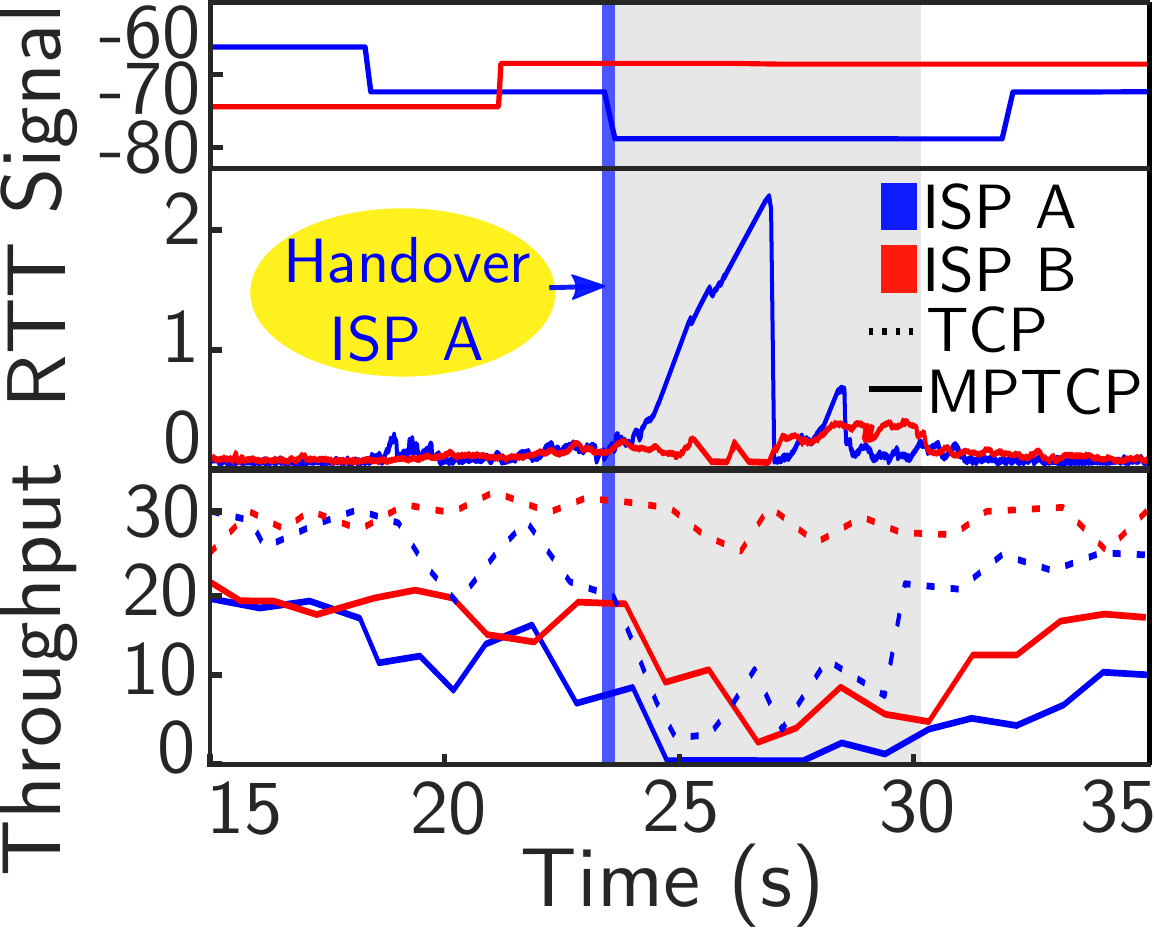}
  \caption{\label{fig:lineHandover}Handover}
\end{subfigure}
\vspace*{-0.5em}
\caption{Effect of network events on throughput (Mbps), RTT (s) of
  MPTCP and parallel TCP. Shaded region denotes RTT recovery time ($t_R$).}
\label{fig:mptcpPhy}
\vspace*{-1em}
\end{figure}

\vspace{-0.5em}
\subsubsection*{\textbf{Impact of network events}} \label{subsec:controlledImpact}
%Here we will talk about the line plots for snr drop and handover. Talk about the recovery time in both cases and observed gap in RTT of both subflows.
%
We now examine the effect of network events on MPTCP performance.
%\noindent \textbf{Impact of Signal Strength} 
%Figure~\ref{fig:lineSNR} shows MPTCP per-flow performance compared to two regular TCP flows while both connections experience signal drops albeit different intensities.
\autoref{fig:mptcpPhy} shows a snippet of throughput over MPTCP and simultaneous TCP over each ISP. \ispa connection observes 8dBm signal drop (\autoref{fig:lineSNR}) and handover (\autoref{fig:lineHandover}).
%
%We also show the RTT observed by both MPTCP subflows but do not display RTT over TCP to maintain graph legibility.
As RTT and signal strength of TCP flows follow a similar behavior as MPTCP, we exclude them from our graph to maintain legibility.
 
We observe from \autoref{fig:lineSNR} that the drop in signal strength on \ispa is immediately followed by $2\times$ RTT
%, $1.7\times$ bytes-in-flight (not shown due to space restriction) 
and results in 22\% decrease in overall throughput.
%We observe that while RSSI drop on \ispa signal reduces regular TCP throughput by $\approx33$\%, its impact on MPTCP subflow is far more significant.  
%We observe that the 4~dB drop in \ispb connection did not impact MPTCP performance as significantly as the 8~dB drop on \ispa signal.
%
%Our explanation to this behavior is as follows.
This result agrees with previous studies which attributes the spike in RTT to increased buffering at the BS as it switches to a lower link rate to accomodate for the drop in client's signal strength~\cite{jiang2012}.
%
%The radio resource management (RRM) at the basestaion switches to a lower link rate on observing a drop in signal strength, thereby resulting in increased buffering.
%
%The additional queue delay on the affected path is corroborated by corresponding spikes in RTT and bytes-in-flight (not shown due to space restriction) indicating the presence of bufferbloat.
%
%Soon after the drop, the established subflow on \ispa experiences RTT spikes as high as \textit{twice} the mean RTT.
%
%We explain this behavior to excessive packet buffering at the base station while the channel switches to a lower bitrate unbeknownst to TCP.
%
%This was corroborated by a corresponding increase in bytes-in-flight at the same time (not shown due to space restriction) indicating the presence of bufferbloat.
%
However, increased RTT impacts MPTCP performance more severely compared to regular TCP (38\%$\downarrow$ vs. 22\%$\downarrow$).
This is due to the behavior of \texttt{minSRTT} scheduler.
%
%As the RTT difference between the subflows increase, 
As \ispa connection observes elevated RTTs, the scheduler opts to send subsequent packets on \ispb to avoid out-of-order deliveries at the receiver.
This continues until the RTT of \ispa improves as the BS flushes out queued packets to the client.
% 
%The shaded region in the figure denotes 
%time taken by MPTCP to recover from after-effects of a network state, \emph{i.e.} 
%time until RTT of both subflows stabilize after network state.
We denote time taken by RTT of affected subflow to recover as $t_R$, shown as the shaded portion of the graph.
%
%The scheduler resumes utilization once the RTT of non-affected subflow exceeds that of the affected. 
%In case of signal drops, this time ranged between 3-6 seconds.
As LTE can experience multiple signal drops while the client moves away from the BS, we find $t_R$ to be as large as 7s and 12s while walking and driving, respectively.
%Upon noticing the RTT increase, MPTCP  sends subsequent packets on the other (\ispb) subflow resulting in much lower utilization of \ispa connection, whereas regular TCP continuously buffers packets until the link rate stabilizes.
%
%We also observe that the time taken by a subflow to recover from queue delays after the signal strength has stabilized far exceeds that of regular TCP. 
%
%We account this behavior to the design of \textit{minSRTT} scheduler  which restricts packets scheduling on affected path until \ispb subflow experiences higher RTTs. 
%
%The presence of  in-network buffers in the core network prevent any packet drops therefore subflow TCP is unable to limit reduce its congestion window size.
%
%Although the affected subflow experiences a throughput hit, MPTCP scheduler schedules packets on ISP B subflow which has much lower RTT.
%
%This allows MPTCP to overcome intermediate path deterioration while maintaining certain throughput.
%Note that the choice of congestion control does not impact MPTCP performance as no TCP packets were reported dropped or re-transmitted.
%none of the connections experienced packets drops leading to high window availability on both flows.

%To summarize, our findings show that drop in signal strength impacts MPTCP more significantly than regular TCP.
%
%Furthermore, the intensity of the drop has a direct correlation to throughput of the subflow.  
%shows that out-of-order queue size at the receiver were as large as 1 MB while walking thereby reducing end-application goodput.

%\noindent \textbf{Impact of Handovers} 
On the other hand, RTT spikes induced by handovers on the subflow surpass those caused by signal drops by a large margin and result in $8\times$ RTT difference between subflows (\autoref{fig:lineHandover}). 
%
%handovers influence RTT of affected subflow far more drastically in contrast to signal drops (\autoref{fig:lineHandover}).
%
% shows that handovers (represented by a colored bar) affect MPTCP performance more severely than regular TCP. 
%
%Our experiment results were found to be in accordance with previous studies on TCP performance in cellular networks \cite{huang2013}as we observed large RTT spikes (upto several seconds )as a result of a handover.
%or one, we observe far larger RTT spikes of intensity upto several seconds on the affected flow which lead to $\approx8\times$ RTT difference between both subflows. 
%
%Our observations are in line with previous studies analyzing TCP behavior over cellular networks~\cite{huang2013}.
%
%Secondly, in contrast to parallel TCP flows and MPTCP's behavior after signal drop events wherein only the affected flow exhibits throughput reduction, both MPTCP subflows are affected following a handover.  
Furthermore, while signal drops cause throughput decline only on the affected subflow, handovers impact the performance of both paths and result in 74\% throughput decrease.
%n the figure, the affected ISP A, observed several acknowledgments arriving as late as 2s!
This trend is absent for TCP flows.
While driving we find that consecutive handovers and signal drops on both links can result in $t_R\geq$40s!
%However, we observe that regular TCP recovers from a handover much quicker than MPTCP primarily due to inefficient path utilization by underlying minSRTT scheduler.
%
Presence of such large delay differences between subflows closely resembles MPTCP behavior over heterogeneous networks \cite{nikraveshmptcpmobile}.
%As a result, MPTCP realizes 74\% decrease in throughput post handover event.
We explain MPTCP's response to handovers in our in-the-wild analysis. 
%To uncover its root cause, we investigate our in-the-wild measurements.
%We demonstrate this behavior in detail later in the paper. 

%We also monitor the out-of-order (OFO) queue size at the receiver and plot its CDF in Figure~\ref{fig:cdfOFO}.
%%
%We observe that sudden drops in signal strength and handovers can result in a large OFO queue buildups at the receiver buffer which often results in head-of-line (HoL) buffer blocking~\cite{blest}. 
%%
%Figure~\ref{fig:cdfOFO} shows that OFO queues can reach $>4$~MB while driving as both connections undergo frequent handovers.
%%
%Such large queue sizes significantly impacts performance on all subflows as the receiver waits for missing packets to free its receive buffer. 

%To learn more about MPTCP reaction to network changes in detail, we focus on uncontrolled results.

\noindent \underline{\texttt{Takeaway2}:} 
\textit{%Write takeaway regarding how MPTCP acts heterogeneously with added mobility. Unlike WiFi+LTE where one network outperforms other for majority, in multi-carrier the performance can be oscillating leading to worse MPTCP adaption.
%There exists a direct correlation between last-mile link changes to drop in MPTCP performance,
%to the extent that handovers impact on throughput exceeds that of a signal strength drop.
Last-mile changes have a direct correlation to MPTCP performance degradation.
% with handovers causing more disruptions compared to signal strength.
%to the extent that handovers impact on throughput exceeds that of a signal strength drop.
%
%While drop in signal strength impacts only affected subflow, handover on one connection exerts influence on both subflows.
%
Following \texttt{Takeaway1}, 
% where one connection constantly outperforms the other, 
%there is no consistently "better" subflow in MPTCP over dual-LTE 
%as both connections experience equally frequent link changes at high speeds.   
%unlike heterogeneous networks e.g. WiFi \& ethernet.
%existing solutions for MPTCP over heterogeneous networks are not applicable in dual-LTE environments due to lack of a consistently "better" performing subflow.
%both subflows suffer equally when the client is mobile resulting in a heterogenous network-like behavior. 
%%
%However, existing heterogeneous-specific solutions are not applicable in dual-LTE due to lack of a consistently "better" path.
both subflows observe spikes in RTT as response to network events, resembling a heterogenous network-like behavior. 
However, existing MPTCP solutions are not applicable in dual-LTE due to the lack of a consistently "better" path.}

\subsection{In-The-Wild Measurements} \label{sec:uncontrolled}

\begin{figure}[!t]
\centering
%\begin{subfigure}{0.32\linewidth}
%\centering
%  \includegraphics[width=\textwidth]{Figs/matlab_plots/mobility_stable}
%  \caption{\label{fig:mobility_stable} Static}
%\end{subfigure}%
%\vspace{0.005\textwidth}
\begin{subfigure}{0.48\linewidth}
\centering
  \includegraphics[width=\textwidth]{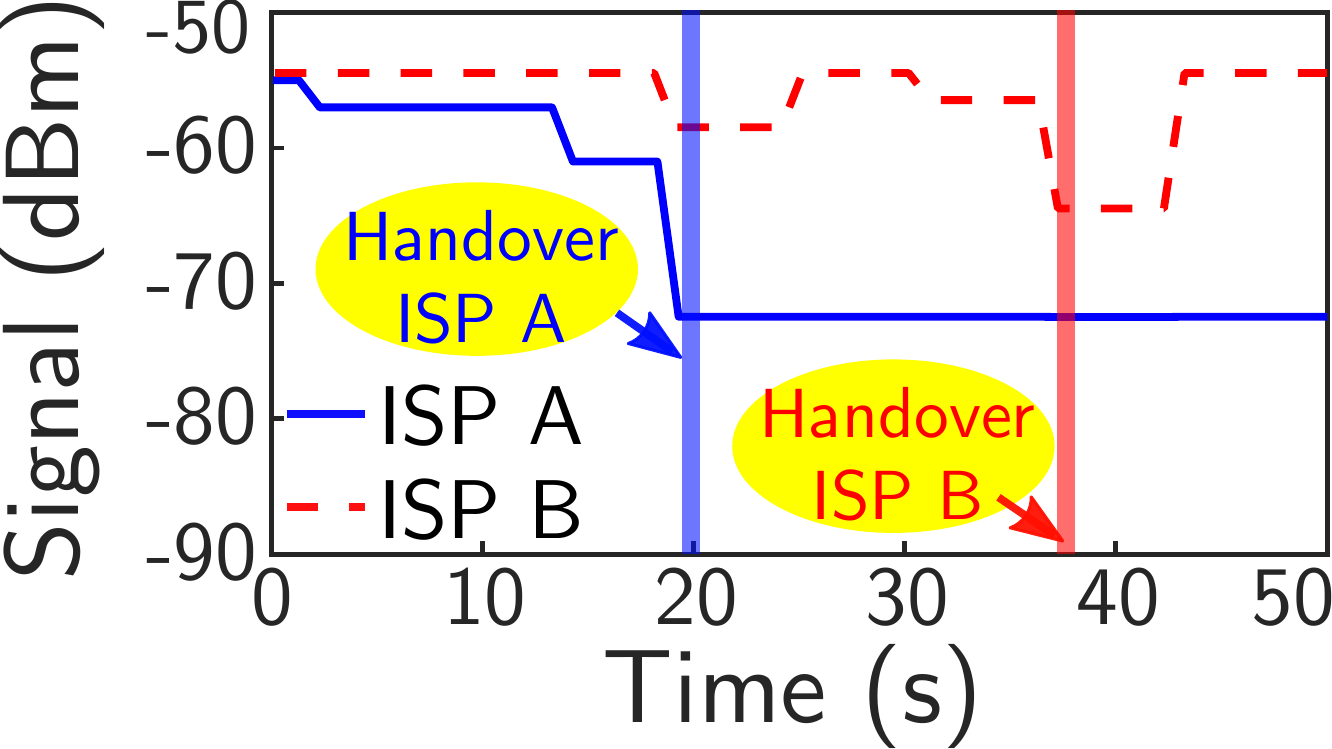}
  \caption{\label{fig:mobility_low}Low mobility}
\end{subfigure}
\hspace{0.005\textwidth}
\begin{subfigure}{0.48\linewidth}
\centering
  \includegraphics[width=\textwidth]{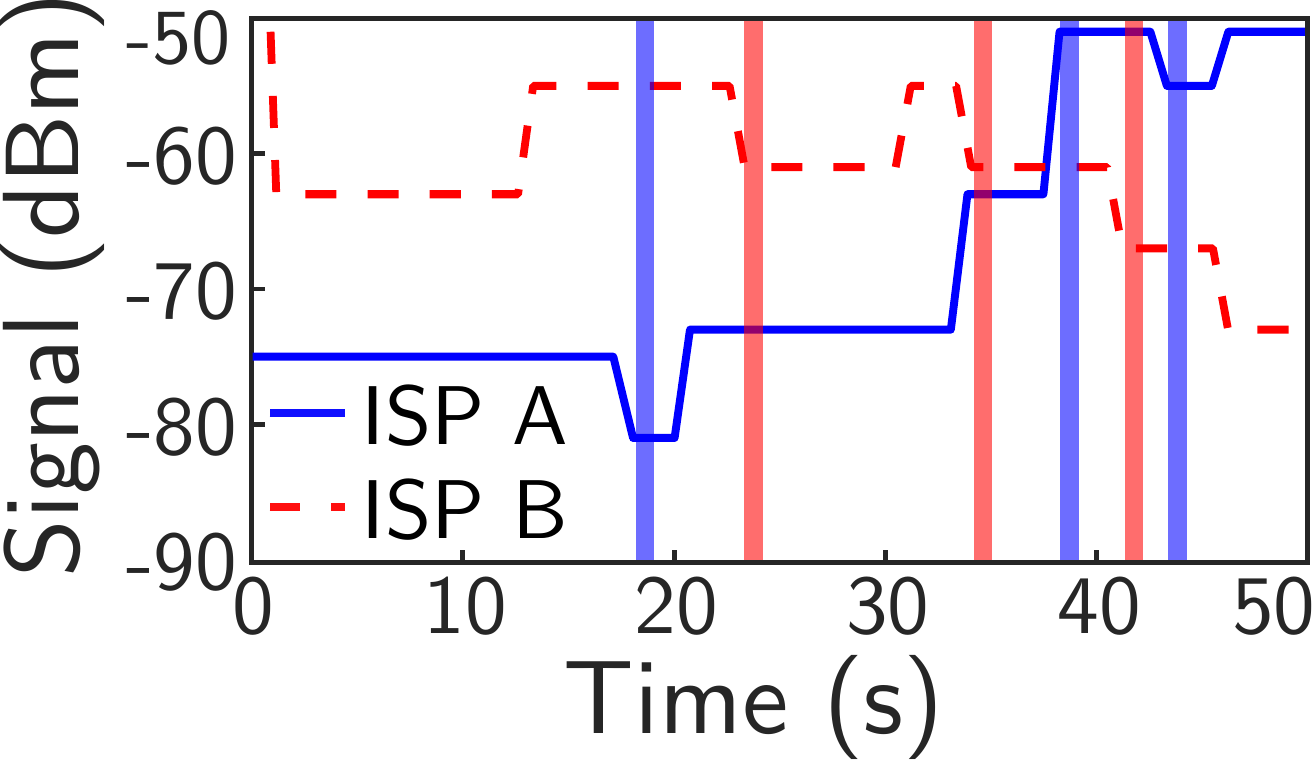}
  \caption{\label{fig:mobility_high}High mobility}
\end{subfigure}
\vspace*{-0.5em}
\caption{Network quality signatures for classifying data captured in in-the-wild into mobility categories.}
\label{fig:mobility_signature}
%\vspace*{-1em}
\end{figure}

Before we analyze MPTCP behavior in-the-wild, we first need to accurately classify the collected data traces into different mobility categories based on volunteer's speed at collection time.
Using our observations in \cref{sec:controlled}, we design a \textbf{classification model} which labels data traces as \emph{low-speed} and \emph{high-speed} depending on the frequency of observed network events. 
The low-speed category represents traces collected while walking and high-speed constitutes all motorized transport modes.
%wherein each LTE connection witnesses multiple signal drops upto 8dBm and few handovers every minute.
%
%On the other hand, the high-speed classifier shows much more frequent network state changes which are dominated by handover and drops more than 8dBm.
%
\autoref{fig:mobility_signature} shows a representation from both categories.  
Our controlled tests revealed that this model achieves 98.7\% classification accuracy. 
%\begin{enumerate}[wide, labelwidth=!, labelindent=0pt]
%\item \textit{Static} covers all data traces where the client remained stationary throughout transfer resulting in no change in network conditions.
%%
%\item 
%%
%This category embodies walking/running mobility.
%%
%\item \textit{High-speed mobility} contains all data traces which observed more than two handovers within two minutes and represents all motorized mobility scenarios.
%\end{enumerate}
%\smallskip

\begin{figure}[!t]
\centering
\begin{subfigure}{0.32\linewidth}
\centering
  \includegraphics[width=\textwidth]{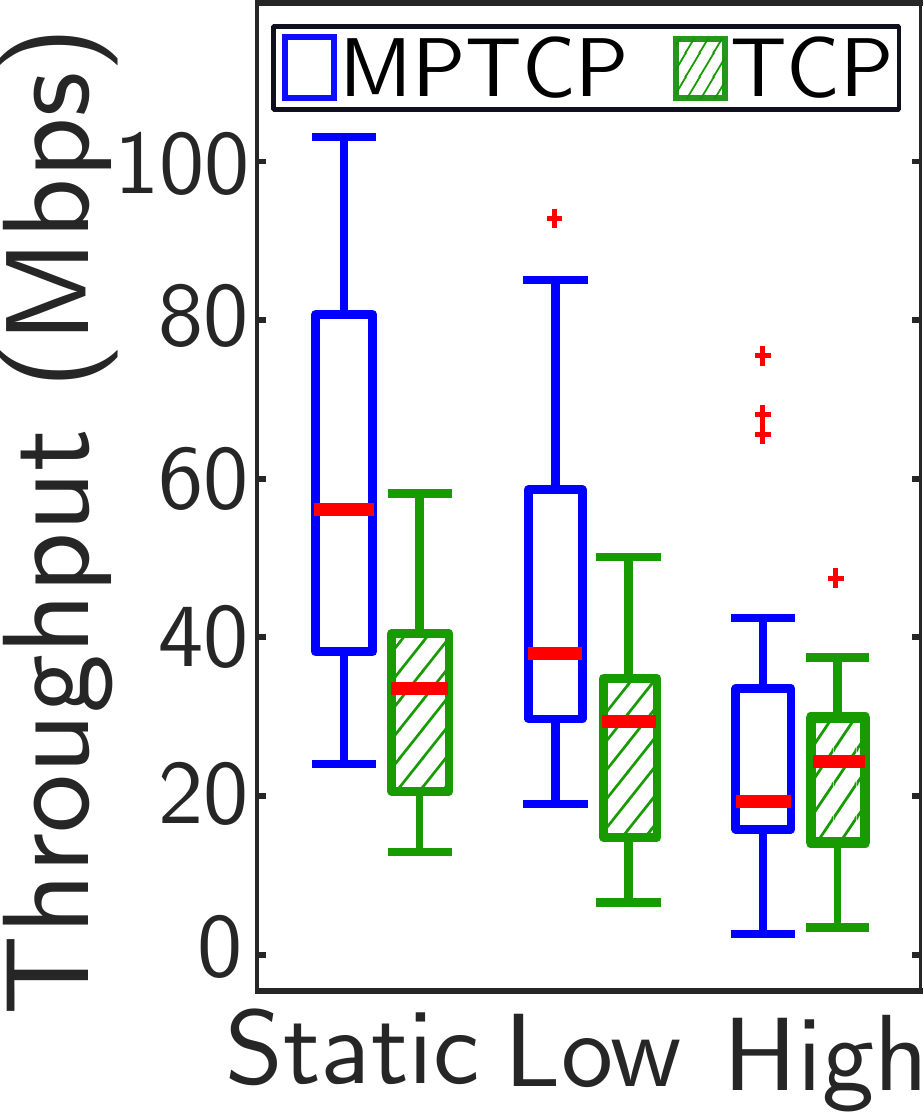}
  \caption{}
  \label{fig:mptcp_tcp_throughput}
\end{subfigure}%
\hspace{0.002\textwidth}
\begin{subfigure}{0.32\linewidth}
\centering
  \includegraphics[width=\textwidth]{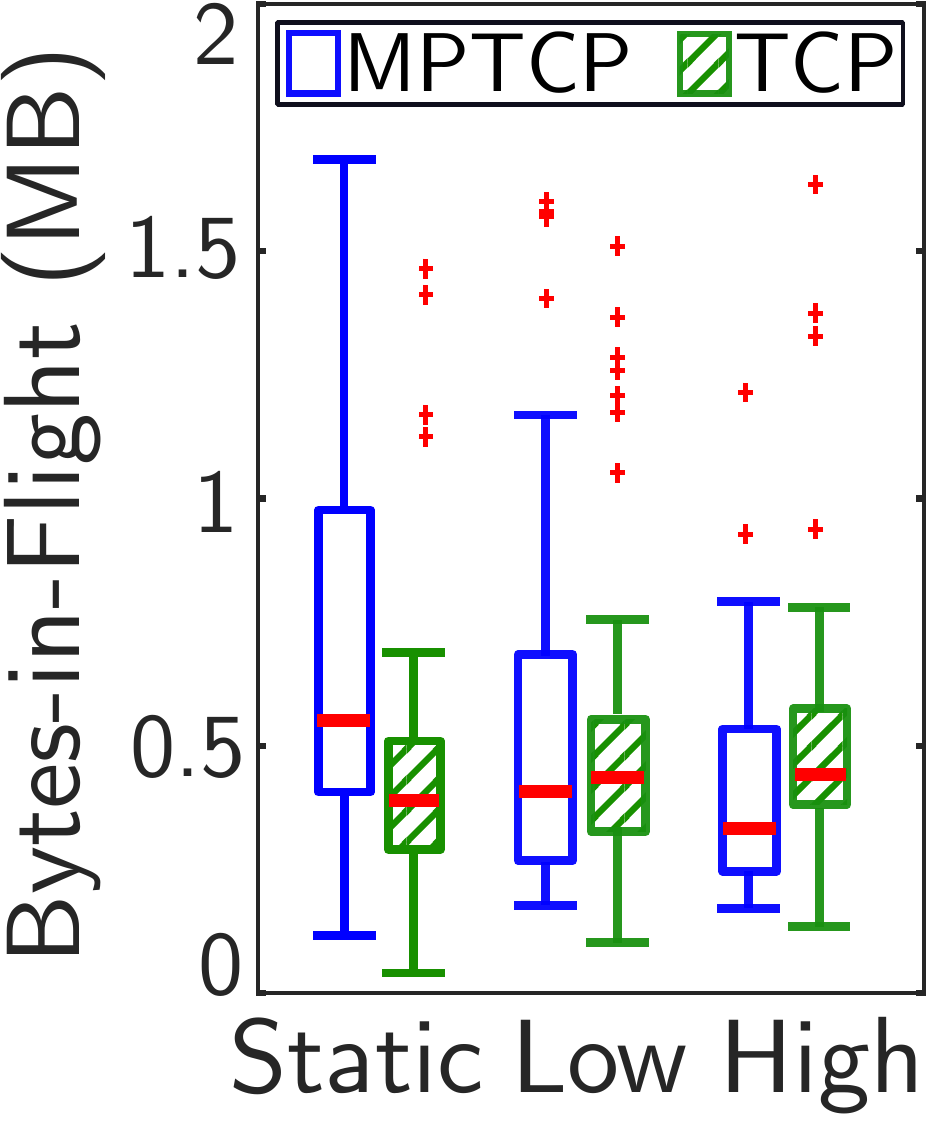}
  \caption{}
  \label{fig:mptcp_tcp_bif}
\end{subfigure}
\hspace{0.002\textwidth}
\begin{subfigure}{0.32\linewidth}
\centering
  \includegraphics[width=\textwidth]{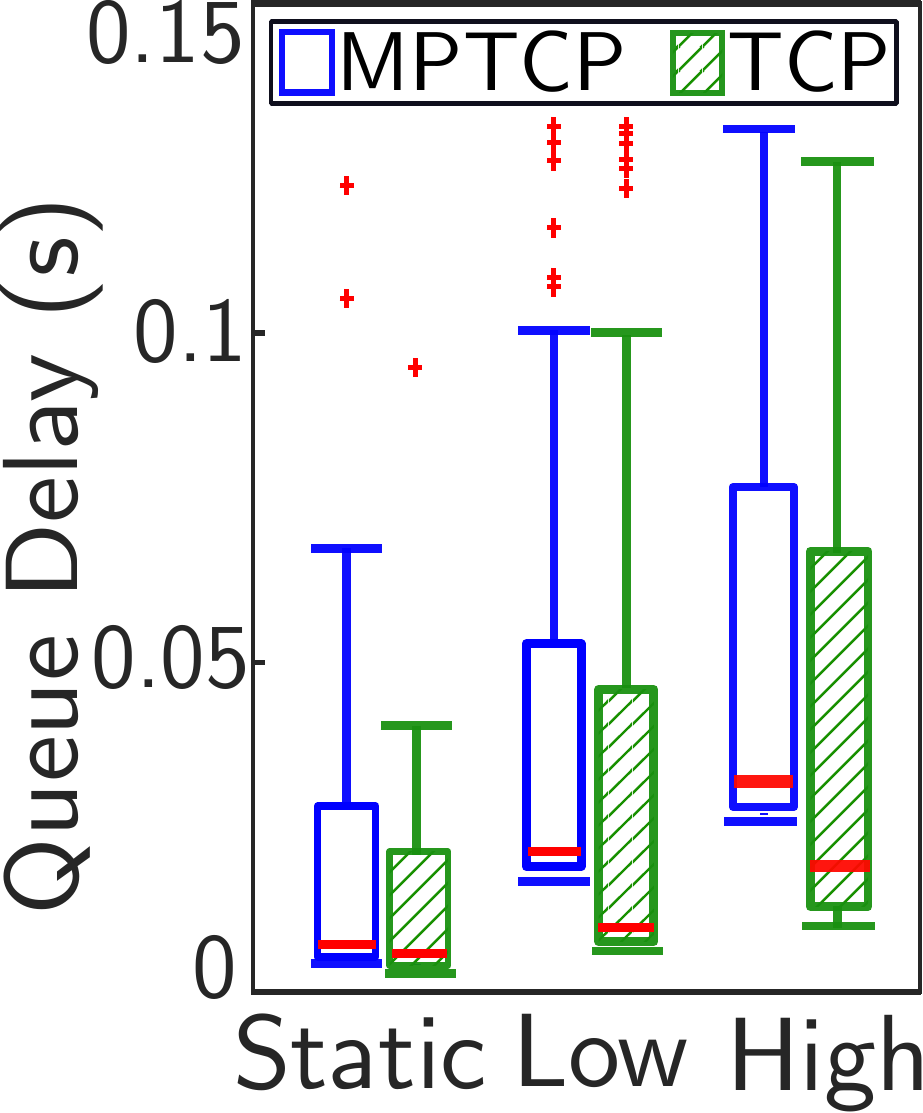}
  \caption{}
  \label{fig:mptcp_tcp_delay}
\end{subfigure}
\vspace{-1\baselineskip}
\caption{Comparison between MPTCP and TCP in (a) throughput achieved, (b) bytes-in-flight and (c) queuing delays for different mobility classes.}
\label{fig:mptcp_tcp_overall}
\vspace*{-1em}
\end{figure}

%\noindent \textbf{How "Multi-Path" is MPTCP?}\
\vspace{-0.5em}
\subsubsection*{\textbf{How "Multi-Path" is MPTCP?}} \label{subsec:wildHowMultipath}
\begin{figure}[!t]
\centering
\begin{subfigure}{0.32\linewidth}
\centering
  \includegraphics[width=\textwidth]{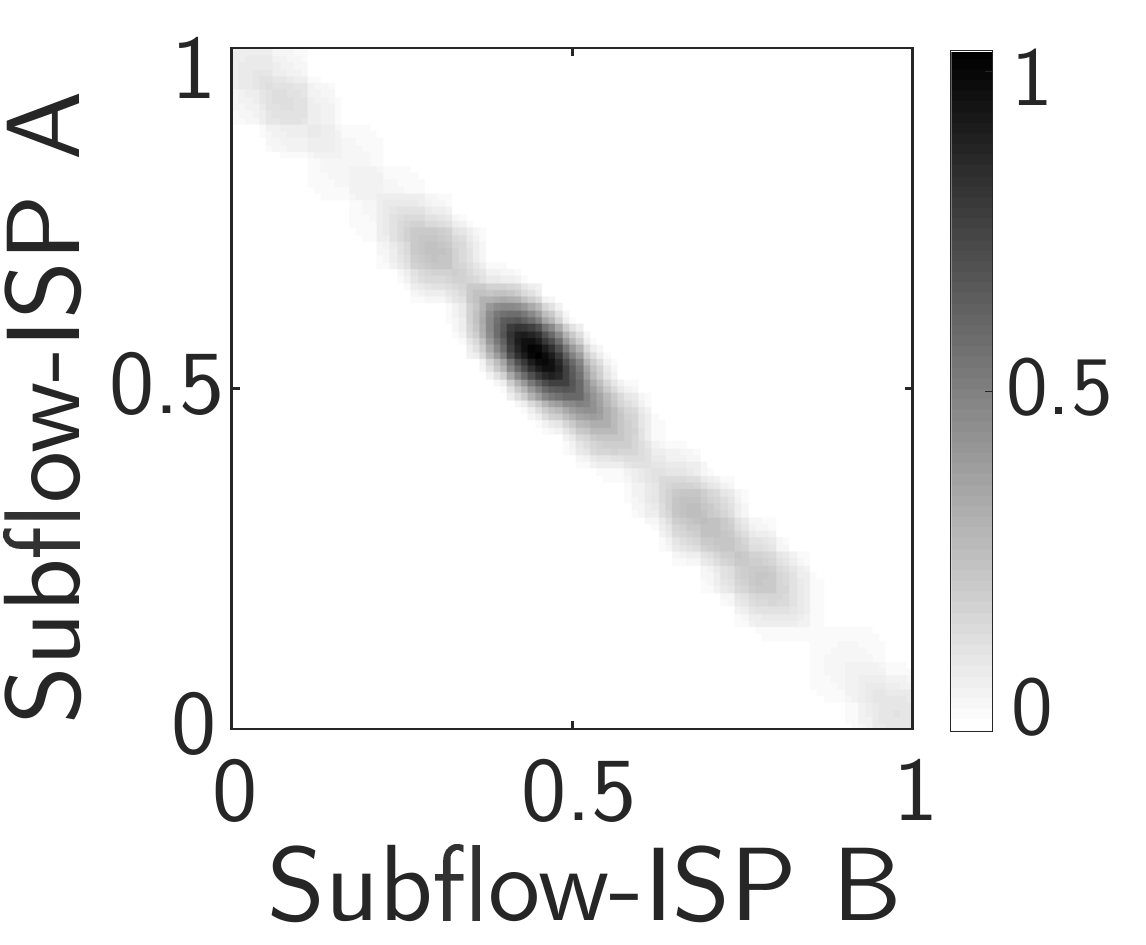}
  \caption{\label{fig:mptcp_heat_stable} Static}
\end{subfigure}%
\hspace{0.001\textwidth}
\begin{subfigure}{0.32\linewidth}
\centering
  \includegraphics[width=\textwidth]{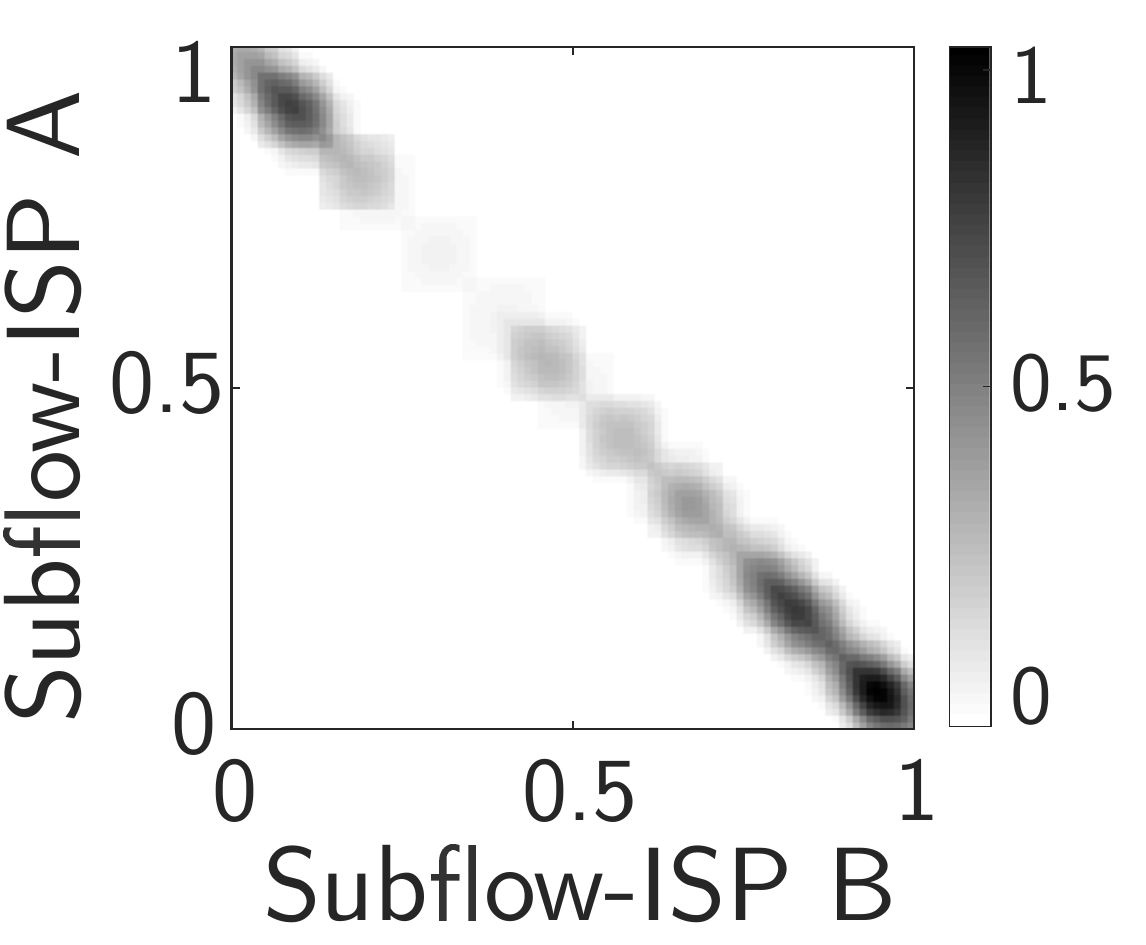}
  \caption{\label{fig:mptcp_heat_low}Low mobility}
\end{subfigure}
\hspace{0.001\textwidth}
\begin{subfigure}{0.32\linewidth}
\centering
  \includegraphics[width=\textwidth]{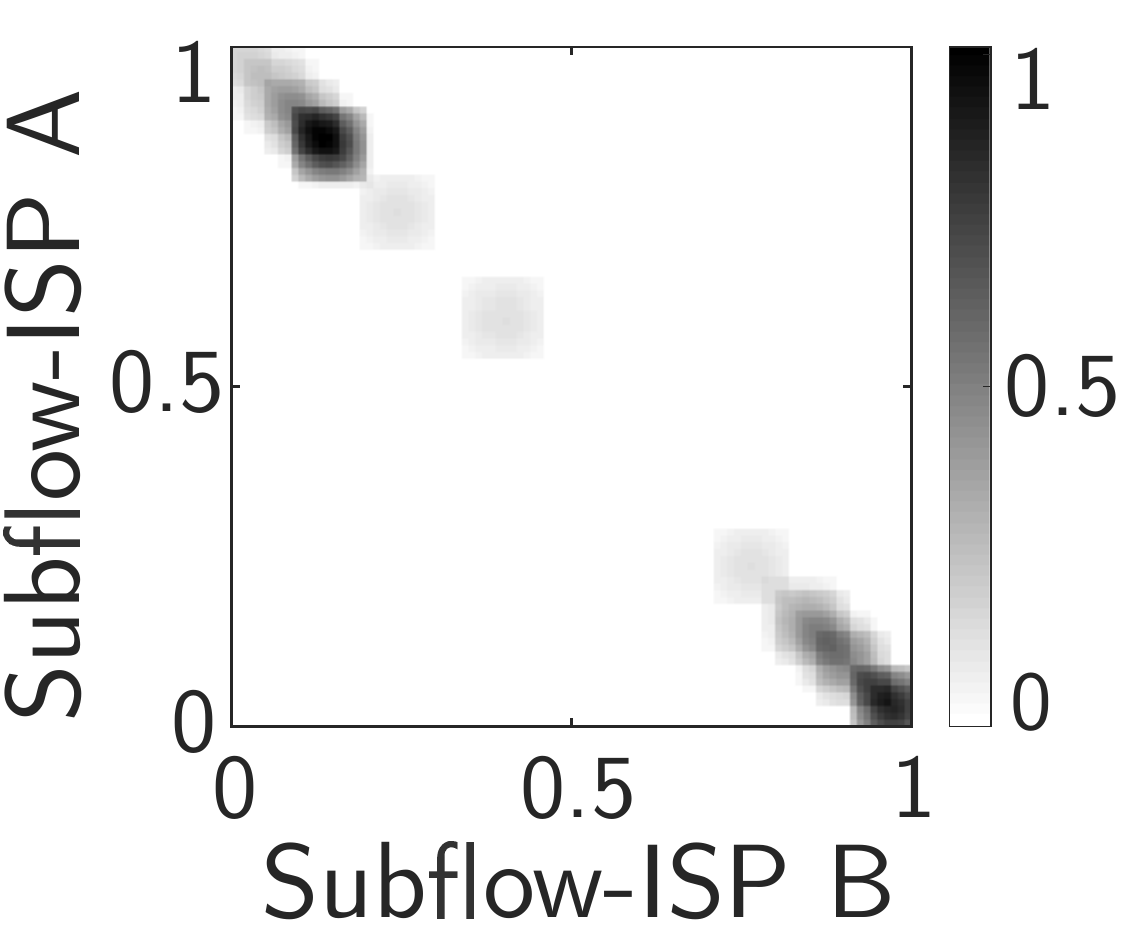}
  \caption{\label{fig:mptcp_heat_high}High mobility}
\end{subfigure}
\caption{MPTCP subflow utilization with increasing mobility. Heat near corners denote skewed usage.}
\label{fig:utilization}
%\vspace*{-1em}
\end{figure}

%
%We first focus on verifying our observations in controlled measurements (\cref{sec:controlled}).  
%We first focus our analysis on understanding advantages of MPTCP over single TCP flow in cellular mobility.
We analyze data traces for trends in network parameters with increasing mobility. %categories for any variations in MPTCP/TCP parameters.  
%
%These results provide finer granularity to our observations from controlled measurements (\autoref{table:control}).
%
\autoref{fig:mptcp_tcp_throughput} shows throughput achieved by MPTCP and TCP over each ISP as a bar plot. 
The results for static and low mobility fall in line with our observations in controlled measurements (\autoref{table:control}).
For high mobility, we find that MPTCP over two LTE connections achieves even lower throughput than TCP over single LTE. 
The result is intriguing as it directly contradicts the basic design
goal of MPTCP to at least perform as-good-as TCP over better path~\cite{mptcpdesign}.
\autoref{fig:utilization} provides us with more insight.
%
%shows the fraction of bandwidth utilization of both MPTCP subflows.
%
For a static client, MPTCP uses both subflows almost equally (heat
concentrated in the center of the diagonal) resulting in $1.6\times$ throughput compared to TCP.  
%We start by analyzing the baseline i.e., static mobility.
%
%MPTCP achieves close to \textit{double} throughput over single TCP flow when the client is stationary.
However, with increasing speeds, the utilization skews towards one of the subflows which triggers 65\% drop in throughput for low-speed mobility.  
%
%Interestingly, we find that TCP over single LTE outperforms MPTCP using two connections at high speeds!
%
In high-speed transportation modes, such as metro, trains etc., MPTCP limits its use to single LTE connection; that too inefficiently as evident by the declining trend in the average number of in-flight packets which seems to be absent for TCP (see \autoref{fig:mptcp_tcp_bif}).
As discussed in \cref{sec:controlled}, the degradation is caused due to \texttt{minSRTT} scheduler's response to RTT spikes induced by network events. 
%
%We now focus on understanding the root cause of such spikes in the first place.
%To find the root cause of such spikes, we plot
%%
%%\autoref{fig:mptcp_tcp_delay} plots 
%the \emph{on-path queuing delay} experienced by MPTCP packets in \autoref{fig:mptcp_tcp_delay} (calculated as instantaneous delay in excess of minimum RTT throughout transfer).
%We investigate the cause of such spikes by plotting the distribution of \emph{queuing delay} endured by MPTCP packets in \autoref{fig:mptcp_tcp_delay} (calculated as instantaneous RTT in excess of minimum RTT throughout transfer).
To find the root cause of such spikes, we plot
%
%\autoref{fig:mptcp_tcp_delay} plots 
the distribution of \emph{queuing delay} endured by MPTCP packets in \autoref{fig:mptcp_tcp_delay} (calculated as instantaneous delay in excess of minimum RTT throughout transfer).
%\autoref{fig:mptcp_tcp_delay} and \autoref{fig:mptcp_tcp_bif} shows the reason.
%We attribute this behavior to fairly equal utilization of both LTE connections by MPTCP (shown as a heatmap in Figure~\ref{fig:mptcp_heat_stable}) as both paths offer similar end-to-end delay and bandwidth to server.
%
Our suspicion of bufferbloat at the basestation is proven correct 
%while the last-mile link stabilizes.
as both MPTCP and TCP experience increasing delays with changes on last-mile.
%Also, as the underlying network conditions remain unchanged throughout the transfer, MPTCP maintains a high send rate and fairly low queue delay (calculated as per-packet delay minus minimum RTT observed) as shown in Figure~\ref{fig:mptcp_tcp_delay_stable} and~\ref{fig:mptcp_tcp_bif_stable}. 
75\% MPTCP packets at high speeds experience queuing delays
as large as $1.7\times$ end-to-end RTT compared to a stationary receiver.

%path utilization

\begin{figure}[!t]
    \centering
    \begin{subfigure}{0.45\linewidth}
        \includegraphics[width=\linewidth]{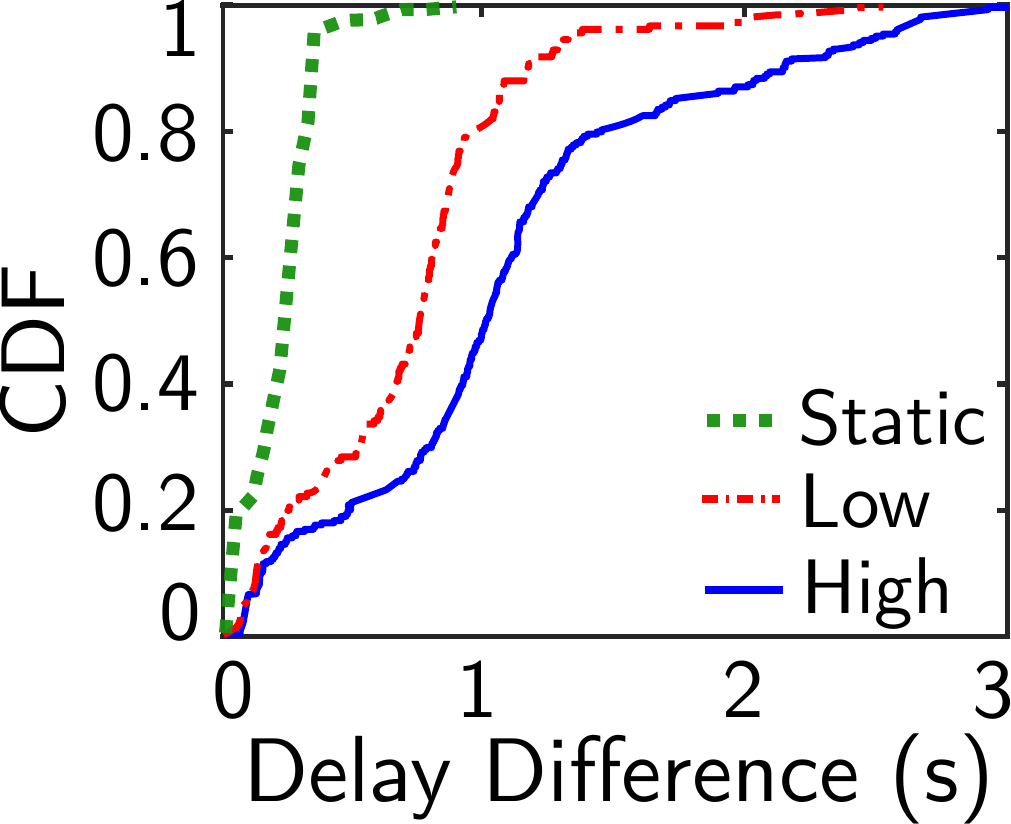} % second figure itself
		\caption{}        
        %\caption{Observed difference in subflow delays in different network conditions.}
	\label{fig:cdfDelayDiff}
    \end{subfigure}
    \hspace{.01\textwidth}
    \begin{subfigure}{0.45\linewidth}
        \includegraphics[width=\linewidth]{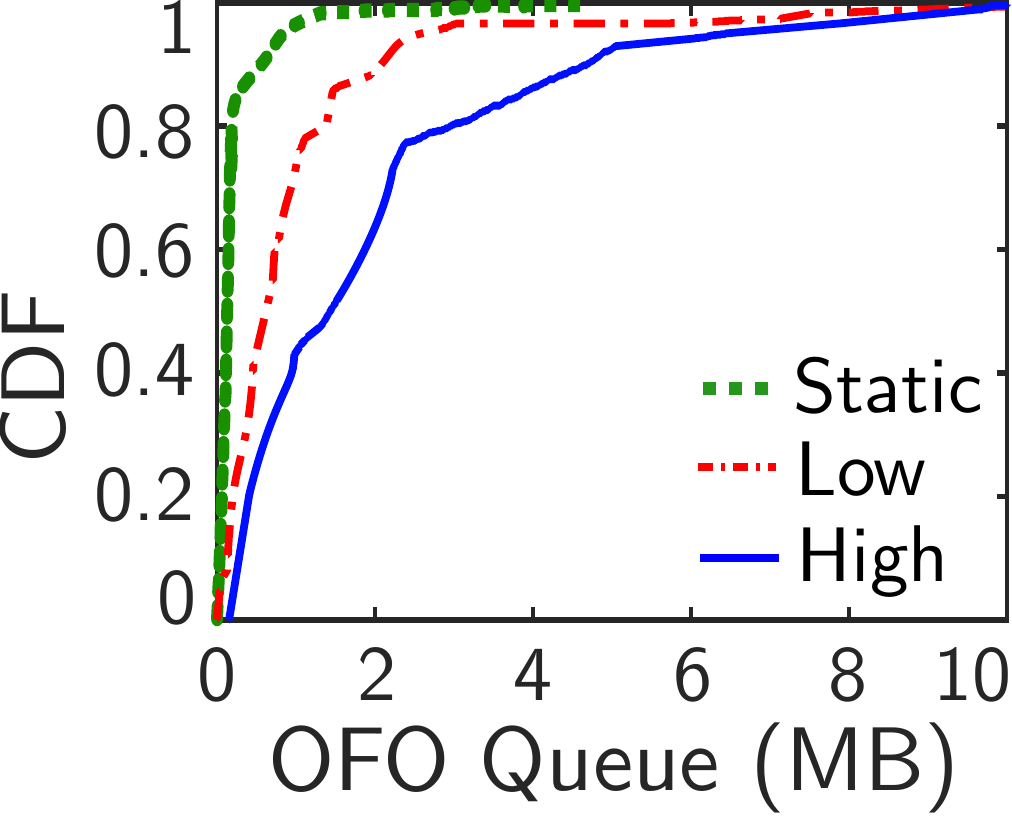}
	\caption{}	
	%\caption{Change in out-of-order queue size with different network conditions.}
	\label{fig:cdfOFO}
    \end{subfigure}
    \vspace{-1\baselineskip}
    \caption{Variation of (a) RTT difference between subflows and (a) out-of-order queue size at receiver.}
\label{fig:overallMPTCP}
\vspace*{-1em}
\end{figure}  

\vspace{-0.5em}
\subsubsection*{\textbf{Dissecting inner-workings of MPTCP}} \label{subsec:wildCause}
%
%We now focus on answering the question, what causes the steep decline in MPTCP's performance at high speed in LTE networks?
%
%From \cref{subsec:controlledImpact}, we note that at higher speeds both LTE connnections encounter simultaneous link disruptions inducing significant RTT differences between both subflows. 
%
We now investigate the consequences of network events on MPTCP decision-making and provide reasons for its behavior to handovers.   
%Although both LTE connections are homogenous, their impact on MPTCP
%due to frequent last-mile changes is akin to MPTCP behavior over
%heteregeneous networks \cite{nikraveshmptcpmobile}. 
%
To this end, we first examine the impact of induced RTT spikes.
\autoref{fig:cdfDelayDiff} shows the distribution of  RTT difference between subflows for different mobility.
75\% MPTCP connections observe 
%>1s inter-subflow delays at low speeds which increases two-fold at higher speeds, almost  
>1s delay gap between both subflows at higher speeds.
%delay in contrast to stationary client.  
%
The primary reason for increasing delays is growing occurrences of network events on alternate LTE links which keeps the underlying network continually unstable. 
Considering \autoref{fig:cdfOFO}, its impact on MPTCP becomes apparent with increased out-of-order buffer occupancy at the receiver.
Larger the occupancy, longer a packet waits in the buffer before being delivered to the application in-ordered sequence.
At high speeds, the out-of-order queue size increases to accommodate packets experiencing considerable delay differences on both paths, until reaching its maximum capacity.
At this stage, the receiver cannot allow for any more packets due to \emph{buffer stalling}.
% until there is space available in the buffer
%     
\emph{This, along with reordering, is the cause for throughput drop on all subflows witnessed in \cref{subsec:controlledImpact}.} 
  
\noindent \texttt{\underline{Takeaway3}:} \textit{Frequent LTE link changes induce large delay differences between subflows which results in unequal subflow utilization, re-ordering delays and buffer stalling, to the extent that single TCP outperforms MPTCP at high speeds.}

\begin{figure}[!t]
    \centering
	\begin{subfigure}{0.46\linewidth}
        \includegraphics[width=\linewidth]{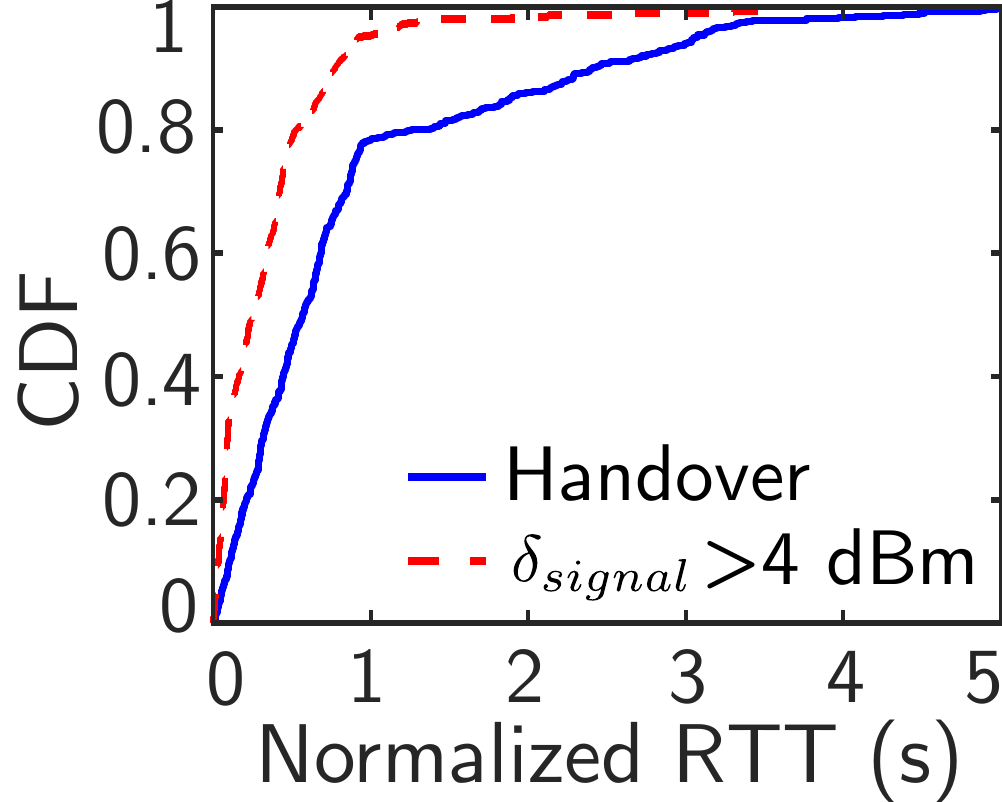} % second figure itself
        \caption{Subflow delay}
	\label{fig:cdfRTT}
    \end{subfigure}    
    \hspace{.01\linewidth}   
    \begin{subfigure}{0.45\linewidth}
        \includegraphics[width=\linewidth]{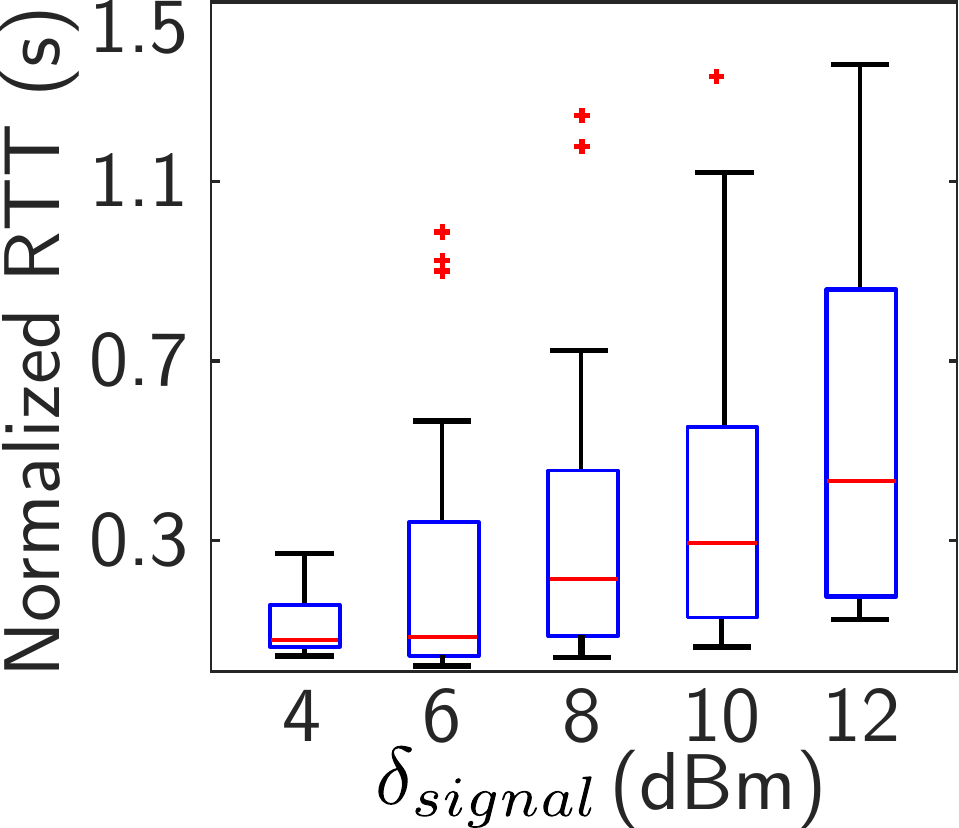} % first figure itself
        \caption{Increasing signal drops}
	\label{fig:snrDropRTT}
    \end{subfigure}
    \caption{Impact of network events on subflow RTT.}
    \label{fig:subflowRTT}
    \vspace*{-1em}
\end{figure}

\vspace{-0.5em}
\subsubsection*{\textbf{Investigating subflow behavior.}} \label{subsec:wildSubflow}
We now explore solutions for improving MPTCP adaptability at high speeds in multi-LTE networks.
%We now provide insights for future solutions to improve MPTCP adaptability at high speeds in LTE.
%    
%In practice, MPTCP allows integration of custom schedulers into existing kernel which can limit subflow utilization based on behavior of network parameters 
%
We begin with investigating the impact of network events on MPTCP subflow,
%, which differs from a regular TCP flow as the rate of packet injections is governed by the MPTCP scheduler instead of the application.
as any trends in subflow performance can be leveraged by MPTCP scheduling policy.
Frequent network events at high speeds, often overlapping, makes 
%accurate estimation of isolated events of a state on subflow  
this analysis challenging.
We carefully separated data traces which had sufficient gaps between consecutive network events.
We calculated \emph{normalized RTT} (instantaneous RTT recorded more than initial RTT at connection establishment) of the affected subflow for associated $t_R$.  
This allows us to identify any spikes in RTT on a subflow post a network event.
\autoref{fig:cdfRTT} shows the distribution of normalized RTT in effect of handover and signal drop on MPTCP subflow.
This distribution validates our results in controlled measurements (\cref{subsec:controlledLastMile}) as it shows $3\times$ and $10\times$ RTT spikes (compared to average RTT) on subflow experiencing signal drop and handover respectively.
%
%\autoref{fig:cdfRTT} shows that 50\% of subflows experience $3\times$ and $10\times$ RTT spikes in contrast to their average RTT (54ms) when subjected to signal drop (186ms) and handover (672ms), respectively. 
%
%Also, the receiver encountered out-of-order deliveries lasting as long as 4 seconds while recovering from handovers as denoted by the 90th percentile.
%
We further dissect the signal drop distribution to analyze the effect of different drop levels on subflow's RTT.
Interestingly, we see a linearly increasing trend emerge indicating that larger signal drops result in higher and longer RTT spikes.
On closer analysis, we find that the trend is deterministic, i.e. \emph{for every 2dB increase in the signal drop, affected subflow observes 1.7-fold RTT spike}.
The result is quite encouraging and suggests that although link changes on last-mile are unpredictable in LTE, their impact on MPTCP can be accurately predicted.

\noindent \texttt{\underline{Takeaway4}:} \textit{A well-designed, cross-layer
MPTCP scheduler, one which actively monitors occurrences of
network events on last-mile can assist in providing robustness and
adaptability over multiple LTE connections.}
%%% Local Variables:
%%% mode: latex
%%% TeX-master: "paper"
%%% End:

%
%\input{controlledExperiments}
%
%\input{uncontrolledExperiments}
%
\section{Discussion} \label{sec:discussion}

% In this section, we discuss the implications of different application loads and effect of congestion control schemes on our results in \cref{sec: results}.

\subsection{Impact of Application Traffic} \label{subsec:video}

\begin{figure}[!tb]
	\centering
	\begin{subfigure}{0.32\linewidth}
		\centering
  		\includegraphics[width=\textwidth]{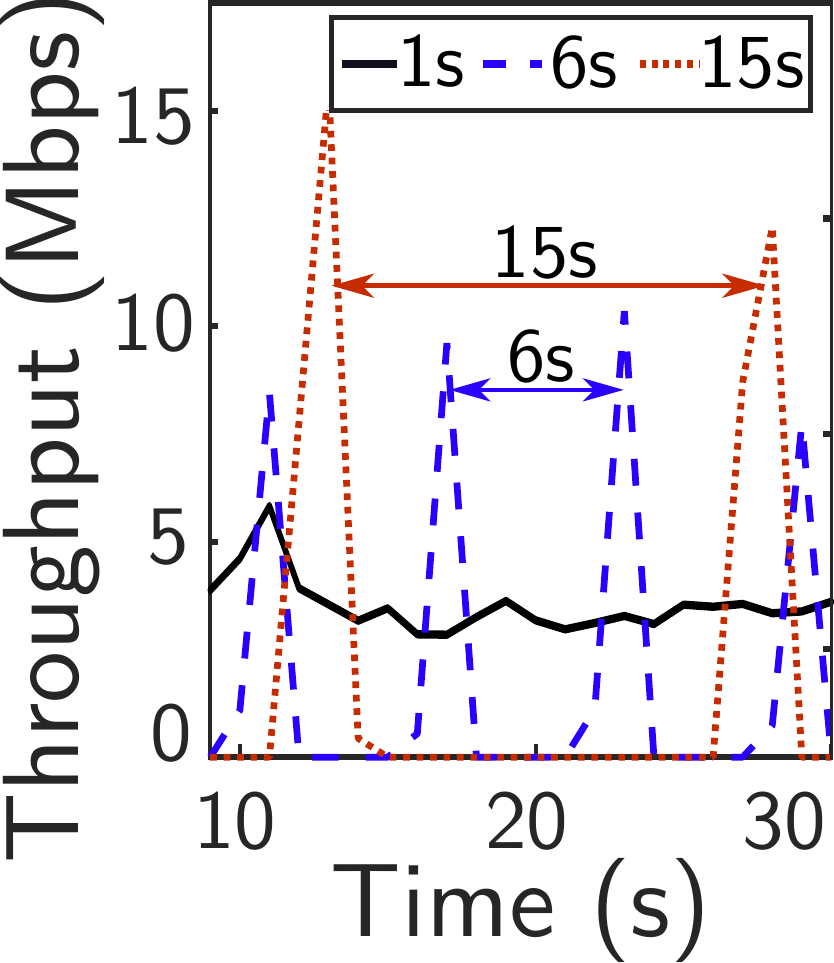}
  		\caption{\label{fig:videopattern} DASH segment traffic pattern}
	\end{subfigure}%
	\hspace{0.002\textwidth}
	\begin{subfigure}{0.32\linewidth}
		\centering
  		\includegraphics[width=\linewidth]{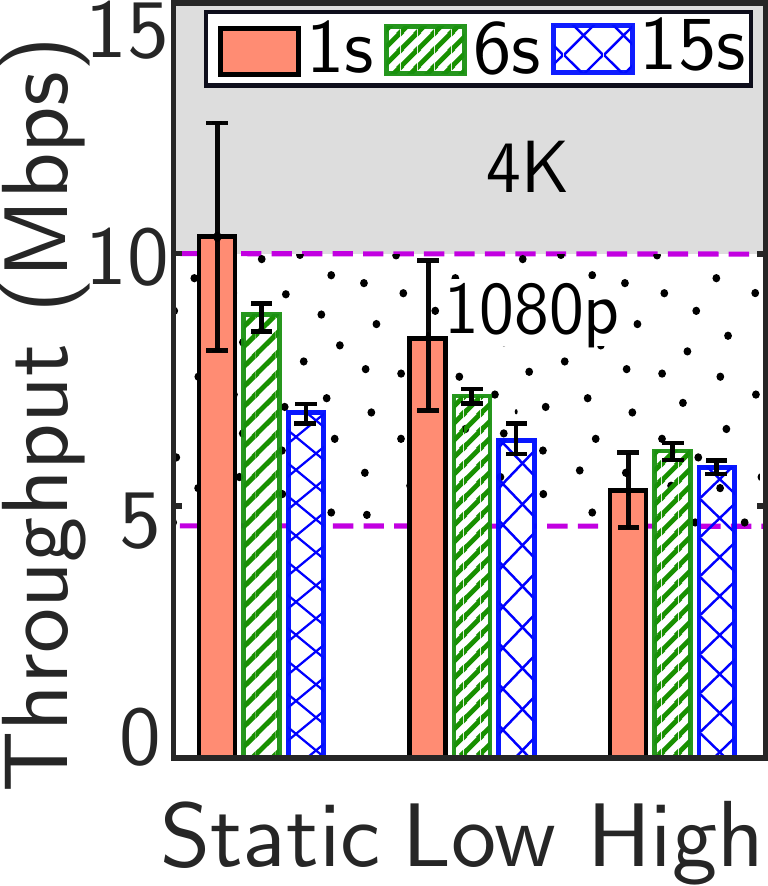}
		\caption{\label{fig:videoThroughput}Throughput (higher is better)}
	\end{subfigure}%
	\hspace{0.002\textwidth}
	\begin{subfigure}{0.32\linewidth}
		\centering
  		\includegraphics[width=\linewidth]{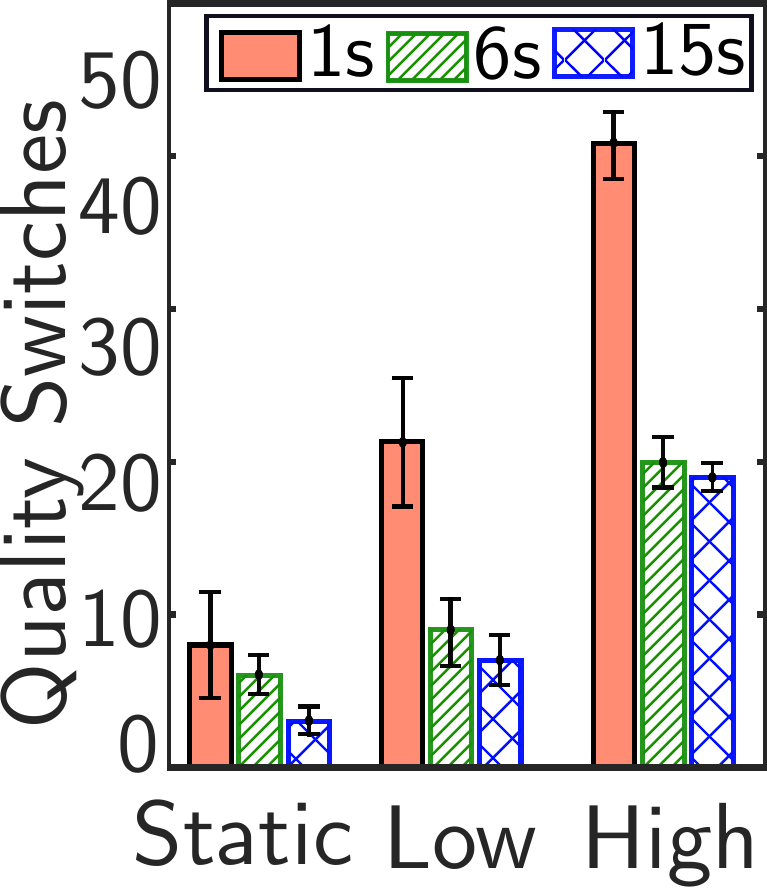}
		\caption{\label{fig:videoRateSwitch}Video QoE (lower is better)}
	\end{subfigure}%	
	\vspace*{-0.5em}	
	\caption{Video streaming over dual-LTE MPTCP.} %(a) shows the traffic pattern for different video segments and (b) shows throughput achieved over both subflows.
\vspace*{-1em}
\end{figure}

%
%\textcolor{red}{Talk about throttling the connection to 7Mbps each to avoid excessive bandwidth availability.}
%
%Previous research has shown that application workload directly impact MPTCP performance~\cite{chen2013imc}.
%`
We also investigated the impact of mobility on an application's QoE
%The next question we ask is, what is the impact of our observations in \cref{sec:uncontrolled} on an application's QoE?
%
by analyzing performance of video streaming over MPTCP for two reasons.
\emph{First}, video streaming accounts for the largest share of mobile traffic in the network~\cite{ciscowhitepaper}.
\emph{Second}, different DASH segment sizes allow us to simulate varying application traffic.
%
%Although past studies have shown that MPTCP \textit{can} enhance streaming performance~\cite{mpdash}, we focus on its effect under mobility.
%
%We modify our experiment setup for this analysis as follows. 
%
We set up a DASH server in our AWS instance and host a 10 minute long \textit{Big Buck Bunny} video on it~\cite{bbbvideo}.
The video was encoded in resolutions ranging from 240p to 4K for bitrates from 50Kbps to 15Mbps.
% on our remote server.
%
%The video is encoded with AVC codec in 1920x1080 format 
% and bitrates ranging from 50kbps to 8Mbps.
We manually throttled both LTE connections to 8Mbps each (total 16Mbps) to remove excess bandwidth.
We re-encoded each resolution into three segments ordered by \emph{increasing burstiness}; 1, 6 and 15 seconds (\autoref{fig:videopattern}).
%
%Figure~\ref{fig:videopattern} shows the streaming pattern for each segment size.
%Our choice of the three segments allow us to study affects of different application loads. 
%
%Video stream encoded with \textit{1-second} segments represents continuous object downloads whereas \textit{15-second} segments best simulate bursty traffic.
%
%We set up a streaming client in our RPi equipped with adaptive bit rate (ABR) which downloads segments of different resolutions depending on currently available network capacity.
The VLC video player in RPi
%uses adaptive bitrate to \
downloads segment sizes which can be best supported by available network capacity. 
%
%We discard first 10 seconds of our results as the player 
%We disable client buffering in order to observe object downloads throughout video length.
%
%Figure~\ref{fig:video} shows our results.
Overall, we analyzed $\approx2000$ traces categorized into three mobility groups.  

\autoref{fig:videoThroughput} shows achieved throughput. Shaded regions denote required throughput for maintaining 4K and 1080p quality.   
%
%It is clear that increasing traffic burstiness leads to a lower throughput as the network remains underutilized for majority of transfer.
%
Interestingly, we find that MPTCP performance in mobility differs for different traffic patterns.
%burstiness of application traffic directly impacts MPTCP performance in increasing cellular mobility.
% where evenly distributed traffic is worst affected by higher speeds}.
%
%As the DASH client initiates segment download nearing the end of previous segment, the number of segments to be downloaded are spread evenly for one second segments.
%
%While streaming 1-second video, frequent network changes causes the client to download lower resolution segments due to instantaneous RTT spikes on either subflow.
% thereby reducing aggregated throughput.
While only constant traffic (1s segment) can support 4K in the static category, it is also affected the worst by high speed and barely achieves 1080p (49\%$\downarrow$ throughput).  
The impact of mobility on its QoE is also substantial as the number of video quality switches while streaming exceeds other segments by 75\% (see \autoref{fig:videoRateSwitch}).
On the other hand, both 6s and 15s streams perform poorly while the client is static, primarily due to limited growth of congestion window size which restricts full bandwidth utilization by bursty traffic.   
However, both segments maintain a consistent 1080p at higher speeds along with minimal quality switches (average 1.7/minute).
%
%We justify this behavior as follows.
%
%On closer analysis, we find that MPTCP congestion window causes this behaviour.
%
%In static conditions the bursty traffic is limited by the slow-start phase of MPTCP congestion window which restricts the application to utilize available bandwidth. 
%
The reason for increased adaptability is a timing mismatch between last-mile changes and application traffic bursts.
%Secondly, while the constant traffic is continously affected by signal drops and handovers at high speeds, its effect in bursty traffic is minimized due to timing mismatch between data spikes and network events on the channel. 
%However, this burstiness assists in bursty streams, the impact of network changes on application is minimized due to timing mismatch between data download and network signal drops or handovers.  
%We notice that with the aggregated bandwidth of two LTE connections MPTCP is able to support bursty 6-second and 15-second streams.
% 
%Even though MPTCP throughput decreases with increase in segment sizes, interestingly, the impact of network change on throughput is also mitigated.
%
%We account this behavior to the bursty traffic load of the application which prioritizes instantaneous large bandwidth over continued availability.
%Furthermore due to large aggregated bandwidth of two LTE connections, MPTCP often/always downloads objects in highest resolution for 6 and 15-second videos respectively despite increasing mobility.
% 
%Furthermore, bursty traffic also has higher probability to avoid data transfers during durations when underlying network is experiencing handovers.
% 
%This result is key in summarizing our experiments, i.e. MPTCP provides instantaneous bandwidth availability under mobility but ineffective utilization over time.

%\noindent \texttt{\underline{Takeaway6}:} \textcolor{red}{Write takeaway}

\subsection{Effect of Congestion Control}
\begin{figure}[!t]
\centering
\begin{subfigure}{0.32\linewidth}
\centering
  \includegraphics[width=\textwidth]{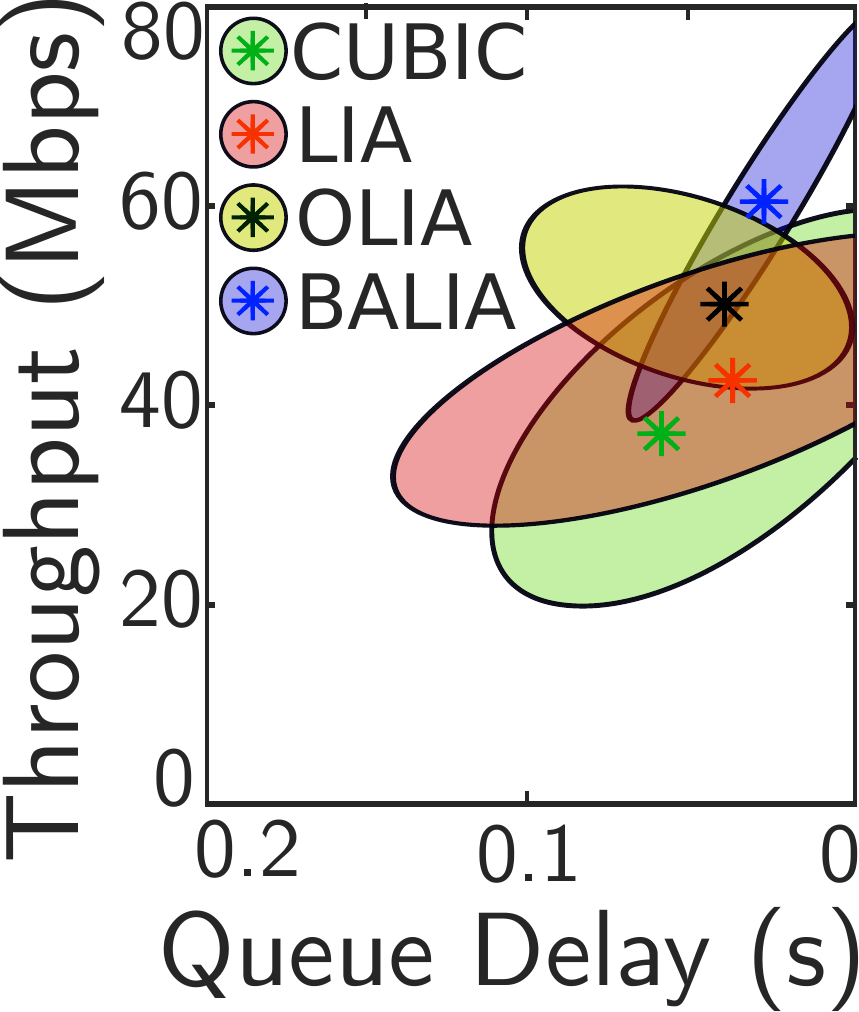}
  %\caption{\label{fig:stableMobility_congestion} Static}
\end{subfigure}%
\hspace{0.003\textwidth}
\begin{subfigure}{0.32\linewidth}
\centering
  \includegraphics[width=\textwidth]{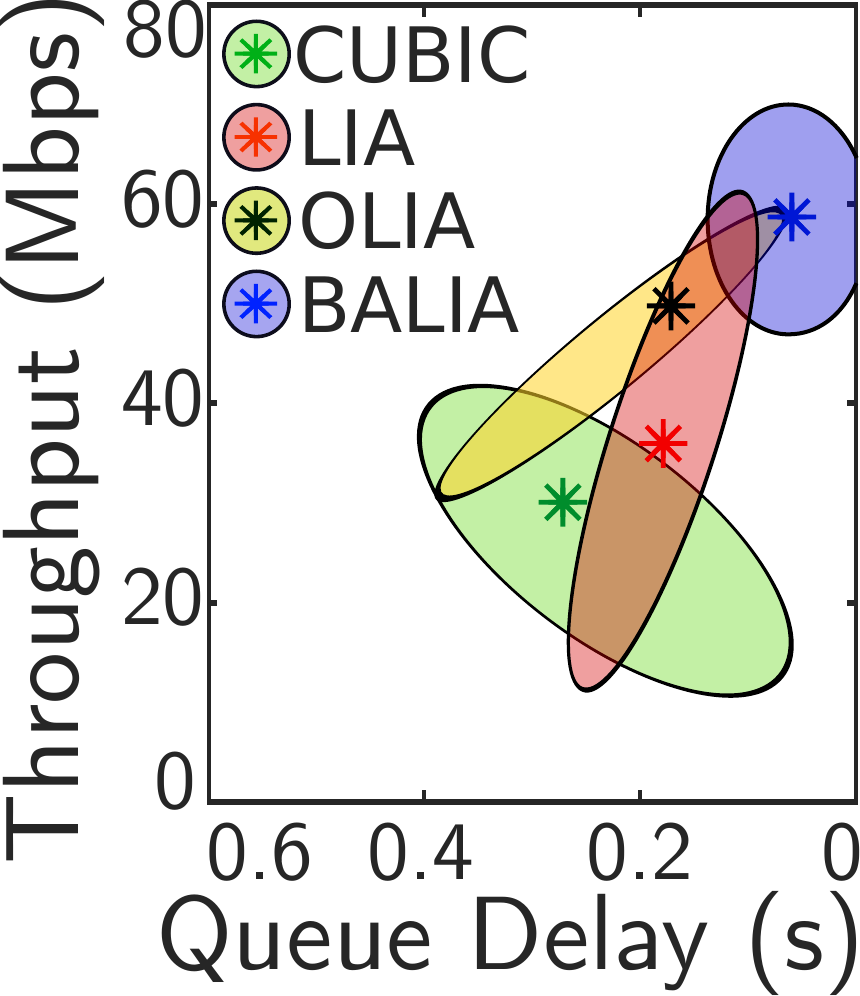}
  %\caption{\label{fig:lowMobility_congestion}Low}
\end{subfigure}
\hspace{0.002\textwidth}
\begin{subfigure}{0.32\linewidth}
\centering
  \includegraphics[width=\textwidth]{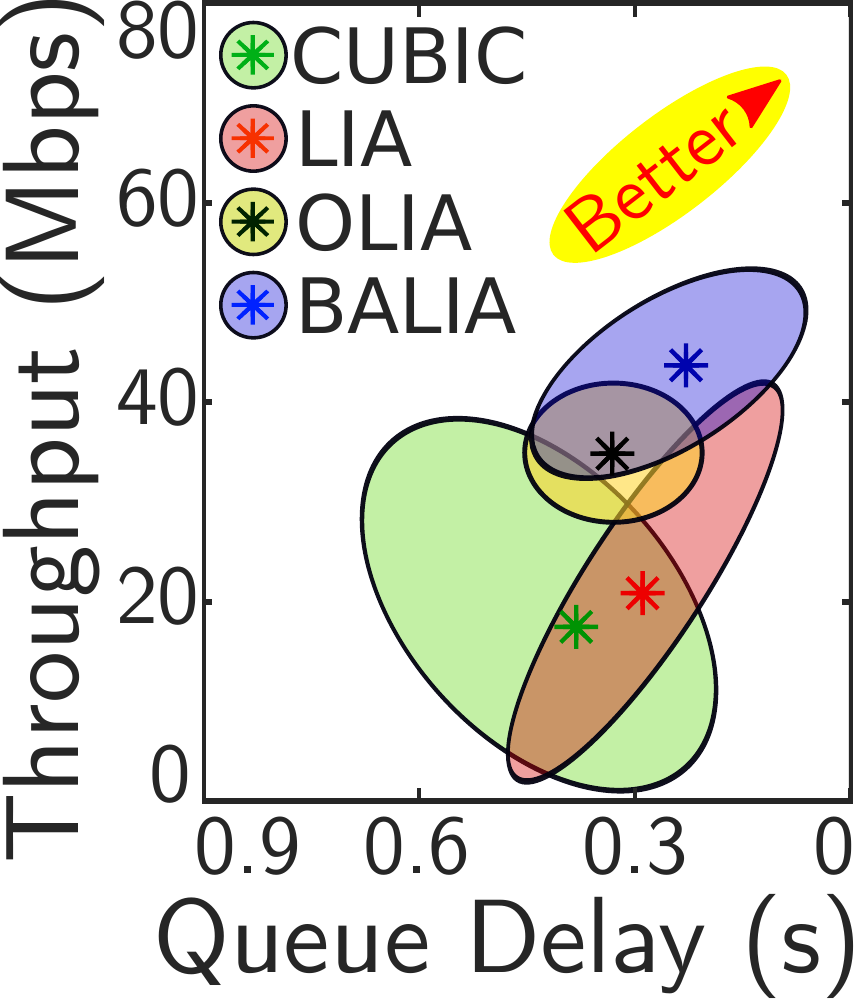}
  %\caption{\label{fig:highMobility_congestion}High}
\end{subfigure}
\caption{Performance of MPTCP congestion control algorithms in (a) static, (b) low and (c) high mobility.}
\label{fig:congestion_mptcp_tcp}
\vspace*{-1em}
\end{figure}
We also explored the impact of different congestion control schemes on cellular mobility.
We conducted a study where we switch the congestion control algorithm in our server and RPi to MPTCP coupled variants, i.e., LIA~\cite{lia}, OLIA~\cite{olia} and BALIA~\cite{balia}, during our in-the-wild measurements. 
We collected $\approx400$ download traces for each scheme and compared it with default uncoupled CUBIC (\cref{sec:uncontrolled}).
\autoref{fig:congestion_mptcp_tcp} presents our results as throughput-delay graphs with flipped x-axis.
%We present our result as throughput-delay graphs shown in Figure~\ref{fig:congestion_mptcp_tcp} where x-axis represents queue delay  flipped to right and y-axis shows throughput achieved. %in Mbps.
%
The graph shows 1-$\sigma$ ellipses for Gaussian distribution of the points. 
An ellipse's orientation signifies covariance between the two axis and asterisks denote median values.
Protocols on the "top-right" are the best on such plots.

%We observe that coupled schemes \textit{do} perform better over uncoupled CUBIC irrespective of mobility.
Coupled schemes out-perform CUBIC while remaining throughput-fair (narrower towards y-axis) with increasing mobility.
Unlike CUBIC, coupled algorithms balance congestion over all subflows and the difference between each variant only lies in their additive increase phase.
% 
%Even though all congestion algorithms show throughput degradation with increasing speeds, uncoupled CUBIC induces largest queue delays.
At higher speeds BALIA outperforms other available flavors, achieving 18\% increase in throughput and 17\% decrease in queue delay. 
OLIA closely follows it and displays similar behavior.
%LIA, by design, employs same increment factor for all subflows and does not allow subflow with smaller congestion window to grow as fast as its peers leading to traffic imbalance. 
%
%Overall out of all coupled protocols, LIA performs the worst followed by OLIA.
%
While we report our observations in this work, 
%we see future studies to be required to dissect these results further. 
a detailed analysis of this behavior is left as future work.  

\noindent \texttt{\underline{Takeaway5}:} 
\textit{%\textcolor{red}{Choice. BALIA is best able to balance congestion away from subflows experiencing network changes and realizes highest throughput and lowest queue delay.}
%Shaping the incoming packet rate to MPTCP scheduler by  
Our results show that both application traffic shape and congestion control flavor impacts MPTCP's ability to adapt to last-mile changes, and should be further explored.
%
%Our evaluation shows that with right choice
%Overall, under mobility, we found that BALIA performs the best closely followed by OLIA and then LIA.
%
%To summarize, the choice of congestion control has a direct impact on MPTCP performance in increasing mobility.
}

\section{Conclusion} \label{sec:conclusion}

This paper studied MPTCP behavior over multi-carrier LTE networks in day-to-day mobility scenarios.
%
%We conducted extensive data collection over five months in controlled and in-the-wild environments. 
Following our extensive data collection over five months,
%Our study focused on analyzing effect of physical network changes, such as signal strength drops and handovers, on MPTCP. 
we observed that MPTCP throughput is severely affected with increasing speeds, often performing even worse than a single TCP connection.  
This is primarily due to frequent last-mile changes on both LTE connections, including signal strength drops and handovers, which result in significant delay differences between MPTCP subflows.
%
%Limited by its receive buffer, 
At high speeds, MPTCP struggled to recover from increased out-of-order transmissions and exhibited a skewed utilization.
%
%However, we discover that the impact of last-mile changes on MPTCP follows a deterministic trend can be solved by well-designed solutions.
We found that effective solutions are possible as the impact of link changes follow a deterministic trend.
%to improve link adaptability at high speeds, tweaking other network parameters can be helpful.  
%
With a better choice of the application traffic pattern and congestion control, MPTCP showed an improvement of 75\% QoE \& 18\% throughput.
%\textcolor{red}{Give some results of the study.}
%We also measured impact of several parameters such as varied application loads, congestion control protocols, etc. in different mobility scenarios. 
%
%Finally, we discussed the possible future work in MPTCP congestion control and scheduler design for effectively adapting to network variability in multi-carrier cellular networks. 
%\textcolor{red}{talk about future work, what should researchers focus on, study and solutions.}

%In our future work, we plan to design a cross-layer scheduler which overcomes delay variability induced by LTE at high speeds.
%%%
%Furthermore, we intend to conduct a future study to understand MPTCP behavior for uploads and the impact of NIC queuing disciplines and existing cross-layer MPTCP schedulers~\cite{shreedhar2018qaware} over LTE at high speeds.  
%%% Local Variables:
%%% mode: latex
%%% TeX-master: "paper"
%%% End:

\bibliographystyle{ACM-Reference-Format}
\bibliography{references}
 
\end{document}